\documentclass[12pt,a4paper]{article}
\usepackage[T1]{fontenc}
\usepackage[sc,osf]{mathpazo}
\usepackage{a4wide}  
\usepackage{amsfonts}
\usepackage{amssymb}
\usepackage{amsmath}
\usepackage{bbm}
\usepackage{booktabs} 
\usepackage{ifpdf}
\ifpdf
\usepackage[pdftex,unicode,implicit]{hyperref}
\hypersetup{%
  pdftitle    = {The tensor hierarchy of 8-dimensional field theories}
  pdfkeywords = {gravity, gauge symmetry, Yang-mills, embedding tensor, tensor hierarchy},
  pdfauthor   = {Oscar Lasso Andino and Tomas Ortin},
  plainpages  = true,
  colorlinks  = true,
  citecolor   = blue,
  urlcolor    = red,
  linkcolor   = black
}
\newcommand{\hepth}[1]{{\tt
\href{http://www.arXiv.org/abs/hep-th/#1}{hep-th/#1}}}

\newcommand{\arxiv}[1]{{\tt
\href{http://www.arXiv.org/abs/#1}{#1}}}
\newcommand{\doi}[1]{{\tt DOI:\href{http://dx.doi.org/#1}{#1}}}
\else
  \usepackage[dvips]{graphicx}
  \usepackage[unicode,implicit]{hyperref}
  \newcommand{\hepth}[1]{{\tt hep-th/#1}}

  \newcommand{\arxiv}[1]{{\tt arXiv:#1}}
  \newcommand{\doi}[1]{{\tt DOI:#1}}
\fi
\makeatletter
\@addtoreset{equation}{section}
\makeatother

\begin{document}

\begin{flushright}
\small
IFT-UAM/CSIC-16-040\\
May 19\textsuperscript{th}, 2016\\
\normalsize
\end{flushright}

\vspace{1.5cm}

\begin{center}

{\Large {\bf The tensor hierarchy of 8-dimensional field theories}}

\vspace{1.5cm}

\renewcommand{\thefootnote}{\alph{footnote}}
{\sl\large  \'Oscar Lasso Andino}\footnote{E-mail: {\tt oscar.lasso [at] estudiante.uam.es}}
{\sl\large and Tom\'{a}s Ort\'{\i}n}\footnote{E-mail: {\tt Tomas.Ortin [at] csic.es}}

\setcounter{footnote}{0}
\renewcommand{\thefootnote}{\arabic{footnote}}

\vspace{1.5cm}

{\it Instituto de F\'{\i}sica Te\'orica UAM/CSIC\\
C/ Nicol\'as Cabrera, 13--15,  C.U.~Cantoblanco, E-28049 Madrid, Spain}\\ \vspace{0.3cm}

\vspace{1.8cm}


{\bf Abstract}

\end{center}

\begin{quotation}
  We construct the tensor hierarchy of generic, bosonic, 8-dimensional field
  theories. We first study the form of the most general 8-dimensional bosonic
  theory with Abelian gauge symmetries only and no massive deformations. This
  study determines the tensors that occur in the Chern-Simons terms of the
  (electric and magnetic) field strengths and the action for the electric
  fields, which we determine. Having constructed the most general Abelian
  theory we study the most general gaugings of its global symmetries and the
  possible massive deformations using the embedding tensor formalism,
  constructing the complete tensor hierarchy using the Bianchi identities. We
  find the explicit form of all the field strengths of the gauged theory up to
  the 6-forms. Finally, we find the equations of motion comparing the Noether
  identities with the identities satisfied by the Bianchi identities
  themselves. We find that some equations of motion are not simply the Bianchi
  identities of the dual fields, but combinations of them.
\end{quotation}

\newpage
\pagestyle{plain}

\tableofcontents

\newpage


\section*{Introduction}


Over the last years, a great effort has been made to explore the most general
field theories. This exploration has been motivated by two main reasons. First
of all there is the need to search for viable candidates to describe the
fundamental interactions known to us (specially gravity) and the universe at
the cosmological scale, solving the theoretical problems encountered by the
theories available today. The second reason is the desire to map the space of
possible theories and the different relations and dualities existing between
them.

In the String Theory context, the landscape of $\mathcal{N}=1,d=4$ vacua has
focused most of the attention, but more general compactifications have also
been studied. At the level of the effective field theories the exploration has
been carried out within the space of supergravity theories. Most ungauged
supergravity theories (excluding those of higher order in curvature) and some
of the gauged ones have been constructed in the past century
\cite{Salam:1989ihk}, but the space of possible gaugings and massive
deformations (related to fluxes, symmetry enhancements etc.~in String theory)
has started to be studied in a systematic way more recently with the
introduction of the embedding-tensor formalism in
Refs.~\cite{Cordaro:1998tx,Nicolai:2000sc,Nicolai:2001sv}. The formalism was
developed in the context of the study of the gauging of $\mathcal{N}=8,d=4$
supergravity in Refs.~\cite{deWit:2002vt,deWit:2005ub}, but it has later been
used in theories with less supersymmetry in different
dimensions.\footnote{See, for instance, Chapter~2 in
  Ref.~\cite{Ortin:2015hya}, which contains a pedagogical introduction to the
  formalism and references.}

The embedding-tensor formalism comes with a bonus: the tensor hierarchy
\cite{deWit:2005hv,deWit:2005ub,deWit:2008ta,Bergshoeff:2009ph,deWit:2009zv}. Using
electric and magnetic vector fields in $d=4$ dimensions as gauge fields
requires the introduction of 2-form-potentials in the theory, which would be
dual to the scalars. In $d=6$ dimensions certain gaugings require the
introduction of magnetic 2-form and 3-form potentials
\cite{Bergshoeff:2007ef}. But the addition of higher-rank potentials does not
stop there: as a general rule, the construction of gauge-invariant field
strengths for the new $p$-form fields requires the introduction of
$(p+1)$-form fields with St\"uckelberg couplings. This leads to a tensor
hierarchy that includes all the electric and magnetic fields of the theory and
opens up the systematic construction of gauged theories: construct the
hierarchy using gauge invariance as a principle expressed through the Bianchi
identities and find the equations of motion by using the duality relations
between electric and magnetic fields of ranks $p$ and $d-p-2$.

This approach has been used in Refs.~\cite{Bergshoeff:2009ph,Hartong:2009vc}
to construct the most general 4-, 5- and 6-dimensional field
theories\footnote{Not only supergravities, since use the embedding-tensor
  formalism is not restricted to supergravity theories.} with gauge invariance
with at most two derivatives. In this paper we want to consider the
8-dimensional case and construct the most general 8-dimensional field theory
with gauge invariance and of second order in derivatives in the action: tensor
hierarchy, Bianchi identities, field strengths, duality relations and
action.\footnote{The tensor hierarchy of maximal 8-dimensional supergravity
  has been constructed in Ref.~\cite{Hohm:2015xna} in the context if
  exceptional field theory.} 

Our main motivation for considering this problem is to simplify and
systematize the construction of a one-parameter family of inequivalent
gaugings with the same SO$(3)$ group of maximal 8-dimensional supergravity,
whose existence was conjectured in
Ref.~\cite{AlonsoAlberca:2000gh}:\footnote{By inequivalent here we mean
  theories which have different interactions, including, in particular,
  different scalar potentials.  A more restrictive definition of inequivalent
  theories (a more general concept of equivalence of theories) is often used
  in the literature (in Ref.~\cite{Dall'Agata:2012bb}, for instance): theories
  related by a field redefinition (including non-local field redefinitions
  such as electric-magnetic dualities) are not considered to be
  inequivalent. With this definition, the theories in the family we are
  talking about would not be considered to be inequivalent.} using
Scherk-Schwarz's generalized dimensional reduction \cite{Scherk:1979zr} Salam
and Sezgin obtained from 11-dimensional reduction an 8-dimensional
SO$(3)$-gauged maximal supergravity in which the 3 Kaluza-Klein vectors played
the role of gauge fields \cite{Salam:1984ft}.\footnote{Other, more general,
  gaugings can be obtained via Scherk-Schwarz reduction
  \cite{AlonsoAlberca:2003jq,Bergshoeff:2003ri}, but it is always the
  Kaluza-Klein vectors that play the role of gauge fields.} The ungauged
theory, though, has a second triplet of vector fields coming from the
reduction of the 11-dimensional 3-form that can also be used as gauge fields
and an SL$(2,\mathbb{R})$ global symmetry that relates these two triplets of
vectors, suggesting one could use as gauge fields any linear combination of
these triplets.

The gauged theory in which the second tripet of vectors (those coming from the
reduction of the 11-dimensional 3-form) played the role of gauge fields was
obtained in Ref.~\cite{AlonsoAlberca:2000gh} by dimensional reduction of a
non-covariant deformation of 11-dimensional supergravity proposed in
Ref.~\cite{Meessen:1998qm,Gheerardyn:2001jj}.\footnote{Many gauged
  supergravities whose 11-dimensional origin is unkonown or, in more modern
  parlance, they contain non-geometrical fluxes (like Roman's 10-dimensional
  massive supergravity or alternative, inequivalent gaugings of other
  theories) can be obtained systematically from this non-covariant deformation
  of 11-dimensional supergravity \cite{AlonsoAlberca:2002tb}, which seems to
  encode many of these non-geometrical fluxes.} This theory has different
Chern-Simons terms and a different scalar potential and provides an early
example of inequivalent gauging with the same gauge group of a given
supergravity theory. However, for the reasons explained above, the existence
of a full 1-parameter family of inequivalent SO$(3)$ gaugings is expected and
it would be interesting to construct it and compare it with the 1-parameter
family of inequivalent\footnote{Inequivalent in the more restrictive sense
  explained above.} SO$(8)$ gaugings of $\mathcal{N}=8,d=4$ supergravity
obtained in Ref.~\cite{Dall'Agata:2012bb} and consider the possible
higher-dimensional origin of the new parameter. 

The construction of that 1-parameter family interpolating between
Salam-Sezgin's theory and that of Ref.~\cite{AlonsoAlberca:2000gh} is a
complicated problem that will be addressed in a forthcoming publication
\cite{kn:OLA}. In this paper we want to consider the general deformations
(gaugings and massive transformations) of generic 8-dimensional theories. This
result paves the way for the constraction of the 1-parameter family of gauged
$\mathcal{N}=2,d=8$ theories which is our ultimate goal. However, it is an
interesting problem by itself whose solution will provide us with the most
general theories with gauge symmetry in 8 dimensions up to two derivatives.

The construction of the most general 8-dimensional theory with gauge symmetry
and at most two derivatives, and this paper, are organized as follows: first,
in Section~\ref{sec-generic}, we study the structure and symmetries (including
electric-magnetic dualities of the 3-form potentials) of generic (up to second
order in derivatives) 8-dimensional theories with Abelian gauge symmetry and
no Chern-Simons terms.

In Section~\ref{sec-abelian}, we consider Abelian, massless deformations of
those theories, which consist, essentially, in the introduction via some
constant ``$d$-tensors'' of Chern-Simons terms in the field strengths and
action. The new intereactions are required to preserve the Abelian gauge
symmetries and, formally, the symplectic structure of the electric-magnetic
duality transformations of the 3-form potentials. We determine explicitly the
form of all the electric and magnetic field strengths up to the 7-form field
strengths, and give the gauge-invariant action in terms of the electric
potentials. This will be our starting point for the next stage. 

In Section~\ref{sec-nonabelian} we consider the most general gauging and
massive deformations (St\"uckelberg couplings) of the Abelian theory
constructed in the previous section using the embedding-tensor formalism. We
proceed as in Refs.~\cite{Hartong:2009vc,Ortin:2015hya}, finding Bianchi
identities for field strengths from the identities satisfied by the Bianchi
identities of the lower-rank field strengths and, then, solving them. We have
found the Bianchi identities satisfied by all the field strengths and we have
managed to find the explicit form of the field strengths up to the 6-form.

In this approach, the ``$d$-tensors'' that define the Chern-Simons terms will
be treated in a different way as in Ref.~\cite{Hartong:2009vc}: they will not
be treated as deformations of the theory to be gauged, but as part of its
definition.  Therefore, we will not associate to them any dual $7$-form
potentials.

In Section~\ref{sec-action} we study the construction of an action for the
theory. The equations of motion are related to the Bianchi identities by the
duality relations between electric and magnetic field strengths, but, at least
in this case, they are not directly equal to them. In general they can be
combinations of the Bianchi identities. To find the right combinations we
derive the Noether identities that the off-shell equations of motion of these
theories should satisfy as a consistency condition that follows from gauge
invariance. Then, we compare those Noether identities with the identities
satisfied by the Bianchi identities. Once the equations of motion have been
determined in this way, we proceed to the construction of the action, which
we achieve up to terms that only contain 1-forms and their derivatives, whose
form is too complicated.

Section~\ref{sec-conclusions} contains our conclusions and the main formulae
(field strengths, Bianchi identities etc.) of the ungauged and gauged
theories are collected in the appendices to simplify their use.


\section{Ungauged  $d=8$ theories}
\label{sec-generic}


In this section we are going to consider the construction of generic
(bosonic) $d=8$ theories coupled to gravity containing terms of second
order or lower in derivatives of any given field\footnote{The
  Chern--Simons (CS) terms may have terms with more than two
  derivatives, but they do not act on the same field.}. The field
content of a generic $d=8$ theory are the metric $g_{\mu\nu}$, scalar
fields $\phi^{x}$, 1-form fields $A^{I}=A^{I}{}_{\mu}dx^{\mu}$, 2-form
fields $B_{m}=\tfrac{1}{2}B_{m\, \mu\nu}dx^{\mu}\wedge dx^{\nu}$ and
3-form fields $C^{a}= \tfrac{1}{3!}C^{a}{}_{\mu\nu\rho}dx^{\mu}\wedge
dx^{\nu}\wedge dx^{\rho}$.  For the moment, we place no restrictions
on the range of the indices labeling these fields nor on the symmetry
groups that may act on them leaving the theory invariant.

We are going to start by the simplest theory one can construct with these
fields to later gauge it and deform it in different ways.

The simplest field strengths one can construct for these fields are their
exterior derivatives:

\begin{equation}
F^{I}\equiv dA^{I}\, ,
\hspace{1cm}
H_{m}\equiv dB_{m}\, ,
\hspace{1cm}
G^{a} \equiv dC^{a}\, .  
\end{equation}

\noindent
They are invariant under the gauge transformations

\begin{equation}
\delta_{\sigma}A^{I} = d\sigma^{I}\, ,
\hspace{1cm}  
\delta_{\sigma}B_{m} = d\sigma_{m}\, ,
\hspace{1cm}  
\delta_{\sigma}C^{a} = d\sigma^{a}\, ,
\end{equation}

\noindent
where the local parameters $\sigma^{I},\sigma_{m},\sigma^{a}$ are,
respectively, 0-, 1-, and 2-forms.

The most general gauge-invariant action which one can write for these fields is 

\begin{equation}
\label{eq:d8simplestgenericaction}
\begin{array}{rcl}
S 
& = & 
{\displaystyle\int} 
\left\{ 
-\star \mathbbm{1} R
+\tfrac{1}{2} \mathcal{G}_{xy}d\phi^{x}\wedge \star d\phi^{y}
+\frac{1}{2}\mathcal{M}_{IJ}F^{I}\wedge \star F^{J}
+\frac{1}{2}\mathcal{M}^{mn} H_{m}\wedge \star H_{n}
\right.
\\
& & \\
& & 
\left.
-\frac{1}{2}\Im\mathfrak{m}\mathcal{N}_{ab} G^{a} \wedge \star G^{b}
-\frac{1}{2}\Re\mathfrak{e}\mathcal{N}_{ab} G^{a} \wedge G^{b}
\right\} \, ,
\\
\end{array}
\end{equation}

\noindent
where the kinetic matrices
$\mathcal{G}_{xy},\mathcal{M}_{IJ},\mathcal{M}^{mn},
\Im\mathfrak{m}\mathcal{N}_{ab}$ as well as the matrix
$\Re\mathfrak{e}\mathcal{N}_{ab}$ are scalar-dependent\footnote{If
  $\Re\mathfrak{e}\mathcal{N}_{ab}$ is constant, then the last term is a total
  derivative.}. One could add CS terms to this action, but this possibility
will arise naturally in what follows.

The equations of motion of the 3-forms $C^{a}$ can be written in the
form\footnote{The equation of motion of a $p$-form field, $\delta S/\delta
  \omega^{(p)}$, is an $(8-p)-form$ defined by  
\begin{equation}
 \delta S 
= 
+\frac{\delta S}{\delta \phi^{x}}\wedge \delta \phi^{x}
+\frac{\delta S}{\delta A_{I}}\wedge \delta A^{I}  
+\frac{\delta S}{\delta B^{m}}\wedge \delta B^{m}
+\frac{\delta S}{\delta C^{a}}\wedge \delta C^{a}\, .
  \end{equation}
With our conventions, when acting on $p$-forms, $\star^{2}=(-1)^{p-1}$.
}

\begin{equation}
\label{eq:Radef}
\frac{\delta S}{\delta C^{a}}
=
-d\frac{\delta S}{\delta G^{a}}
=
0\, ,
\hspace{1cm}
\frac{\delta S}{\delta G^{a}} 
=
R_{a}
\equiv  
-\Re\mathfrak{e}\mathcal{N}_{ab}  G^{b}
-\Im\mathfrak{m}\mathcal{N}_{ab} \star G^{b}\, .
\end{equation}

These equations can be solved locally by introducing a set of dual 3-forms
$C_{a}$ implicitly defined through their field strengths $G_{a}$

\begin{equation}
\label{eq:dualGa}
R_{a}
=
G_{a}
\equiv
dC_{a}\, .
\end{equation}

\noindent
It is convenient to construct vectors containing the fundamental and dual
3-forms:

\begin{equation}
(C^{i}) 
\equiv  
\left(
  \begin{array}{c}
   C^{a} \\ C_{a} \\ 
  \end{array}
\right)\, ,
\hspace{1cm}
G^{i}
\equiv
dC^{i}\, ,
\end{equation}

\noindent
so that the equations of motion and the Bianchi identities for the fundamental
field strengths take the simple form

\begin{equation}
\label{eq:dGi=0}
dG^{i} = 0\, .  
\end{equation}

In other words: we have traded an equation of motion by a Bianchi identity and
a duality relation. In what follows we will do the same for all the fields in
the action so that, in the end, we will have only a set of Bianchi identities
and a set of duality relations between magnetic and electric fields.

The vector of field strengths $G^{i}$ satisfies the following \textit{linear,
  twisted, self-duality constraint}

\begin{equation}
\label{eq:lineartwistedselfdualityconstraint}
\star G^{i} = \Omega^{ij}\mathcal{W}_{jk}G^{k}\, ,
\end{equation}

\noindent
where 

\begin{equation}
\label{eq:omega}
( \Omega_{ij}) 
= 
(\Omega^{ij}) 
\equiv  
  \left(
  \begin{array}{cc}
   0 & \mathbbm{1} \\
-\mathbbm{1} & 0 \\ 
  \end{array}
\right)\, ,
\end{equation}

\noindent
is the symplectic \textit{metric} and 

\begin{equation}
(\mathcal{W}_{ij}(\mathcal{N}))
\equiv 
-
\left(
  \begin{array}{cc}
I_{ab} +R_{ac}I^{cd}R_{db}\,\,\,\,
& 
R_{ac}I^{cb} 
\\
& \\
I^{ac}R_{cb} 
& 
I^{ab}
\\
\end{array}
\right)\, ,
\hspace{1cm}
\Omega \mathcal{W} \Omega^{T} = \mathcal{W}^{-1}\, ,
\end{equation}

\noindent
is a symplectic symmetric matrix\footnote{Basically the same that occurs in
  $d=4$ theories, $\mathcal{M}(\mathcal{N})$ see
  \textit{e.g.}~Ref.\cite{Andrianopoli:1996cm}. We use a slightly different
  convention for the sake of convenience and
  $\mathcal{M}(\mathcal{N})=\mathcal{M}(-\mathcal{N})$ due to the
  unconventional sign on the definition of $G_{a}$.}.  The equations
(\ref{eq:dGi=0}) are formally invariant under arbitrary
$\mathrm{GL}(2n_{3},\mathbb{R})$ transformations ($n_{3}$ being the number of
fundamental 3-forms) but, just as it happens for 1-forms in $d=4$, the
self-duality constraint Eq.~(\ref{eq:lineartwistedselfdualityconstraint}) is
only preserved by $\mathrm{Sp}(2n_{3},\mathbb{R})$. As usual, the only
$\mathrm{Sp}(2n_{3},\mathbb{R})$ transformations which are true symmetries of
the equations of motion are those associated to the transformations of the
scalars which are isometries of $\mathcal{G}_{xy}$ and which also induce
linear transformations of the other kinetic matrices. We will discuss this
point in more detail later on.

The dualization of the other fields does not lead to any further
restrictions. 

In what follows we are going to generalize the simple Abelian theory that we
have constructed by \textit{deforming} it, adding new couplings. We will use
two guiding principles: preservation of gauge symmetry (even if it needs to be
deformed as well) and preservation of the \textit{formal} symplectic
invariance that we have just discussed.


\section{Abelian, massless deformations}
\label{sec-abelian}


The deformations that we are going to consider in this section consist,
essentially, in the introduction of CS terms in the field strengths and in the
action. St\"uckelberg coupling will be considered later. Only the 3- and
4-form field strengths admit these massless Abelian deformations. It is
convenient to start by considering this simple modification of
$G^{a}$:\footnote{We will often suppress the wedge product symbols $\wedge$ in
  order to simplify the expressions that involve differential forms.}

\begin{equation}
\label{eq:Gadeformed1}
G^{a} = dC^{a} +d^{a}{}_{I}{}^{m}F^{I}B_{m}\, ,  
\end{equation}

\noindent
where $d^{a}{}_{I}{}^{m}$ is a constant tensor. The gauge transformations need
to be deformed accordingly:

\begin{equation}
\delta_{\sigma}A^{I} = d\sigma^{I}\, ,
\hspace{1cm}  
\delta_{\sigma}B_{m} = d\sigma_{m}\, ,
\hspace{1cm}  
\delta_{\sigma}C^{a} = d\sigma^{a}-d^{a}{}_{I}{}^{m}F^{I}\sigma_{m}\, .
\end{equation}

\noindent
The action Eq.~(\ref{eq:d8simplestgenericaction}) remains gauge-invariant but
the formal symplectic invariance is broken: if we do not modify the action,
the dual 4-form field strengths are just $G_{a}=dC_{a}$ and
$\mathrm{Sp}(2n_{3},\mathbb{R})$ cannot rotate these into $G^{a}$ in
Eq.~(\ref{eq:Gadeformed1}). Furthermore, the 1-form and 2-form equations of
motion do not have a symplectic-invariant form. 

This problem can be solved by adding a CS term to the action:

\begin{equation}
\label{eq:CS1}
S_{CS}
= 
\int 
\{ -d_{aI}{}^{m}dC^{a}F^{I}B_{m}
\}\, ,  
\end{equation}

\noindent
that modifies the equations of motion of the 3-forms

\begin{equation}
-d\frac{\delta S}{\delta dC^{a}}
=
0\, ,
\hspace{1cm}
\frac{\delta S}{\delta dC^{a}} 
= 
R_{a}-d_{aI}{}^{m}F^{I}B_{m}\, .
\end{equation}

\noindent
The local solution is now

\begin{equation}
dC_{a}
\equiv
R_{a}-d_{aI}{}^{m}F^{I}B_{m}\, ,  
\end{equation}

\noindent
and, since $R_{a}$ is gauge-invariant, the dual, gauge-invariant, field
strength must be defined by

\begin{equation}
R_{a}
=
 dC_{a} +d_{aI}{}^{m}F^{I}B_{m}
\equiv
G_{a}\, .
\end{equation}

Again, $
(C^{i})
=
\left(
\begin{smallmatrix}
  C^{a}\\ C_{a} \\
\end{smallmatrix}
\right)
$
transforms linearly as a symplectic vector if 
$
(d^{i}{}_{I}{}^{m})
\equiv
\left(
\begin{smallmatrix}
  d^{a}{}_{I}{}^{m} \\ d_{aI}{}^{m} \\
\end{smallmatrix}
\right)
$
also does. Then, we can define the symplectic vector of 4-form field strengths

\begin{equation}
\label{eq:Gideformed1}
G^{i} = dC^{i} +d^{i}{}_{I}{}^{m}F^{I}B_{m}\, ,
\end{equation}

\noindent
invariant under the deformed gauge transformations

\begin{equation}
\delta_{\sigma}A^{I} = d\sigma^{I}\, ,
\hspace{1cm}  
\delta_{\sigma}B_{m} = d\sigma_{m}\, ,
\hspace{1cm}  
\delta_{\sigma}C^{i} = d\sigma^{i}-d^{i}{}_{I}{}^{m}F^{I}\sigma_{m}\, .
\end{equation}

However, the deformed gauge transformations do not leave invariant the CS term
Eq.~(\ref{eq:CS1}). The only solution\footnote{We have not found any other.}
is to add another term of the form\footnote{We use the compact notation
  $A^{IJ\ldots}=A^{I}A^{J} \cdots$, $F^{IJ\ldots}=F^{I} F^{J} \cdots$,
  $B_{mn\ldots}=B_{m}B_{n}\cdots$ etc., where we have suppressed the wedge
  product symbols.}

\begin{equation}
\label{eq:CS2}
S_{CS}
= 
\int 
\{ 
-d_{aI}{}^{m}dC^{a}F^{I}B_{m} 
-\tfrac{1}{2}d_{aI}{}^{m}d^{a}{}_{J}{}^{m}F^{IJ}B_{mn}
\}\, ,  
\end{equation}

\noindent
provided the following constraint holds:

\begin{equation}
d_{a(I}{}^{[m}d^{a}{}_{J)}{}^{n]}=0\, ,
\,\,\,\,\,
\mbox{so}  
\,\,\,\,\,
d_{i(I}{}^{(m}d^{i}{}_{J)}{}^{n)}=0\, .
\end{equation}

Observe that we are just using \textit{formal} symplectic invariance: the
symplectic vector $d^{i}{}_{I}{}^{m}$ is \textit{transformed} into a different
one. Thus, in general, one gets $\mathrm{Sp}(2n_{3},\mathbb{R})$ multiplets of
theories, except when $d^{i}{}_{I}{}^{m}$ is a symplectic invariant
\textit{tensor},\footnote{The only symplectic-invariant vector is $0$.} which
requires, at least, one of the indices $I$ or $m$ to be a symplectic index. In
most cases the part of the symmetry group of the theory acting on the 3-forms,
while embedded in $\mathrm{Sp}(2n_{3},\mathbb{R})$, will be a much smaller
group and, then, full symplectic invariance of $d^{i}{}_{I}{}^{m}$ may not be
required.

As a nice check of the formal symplectic invariance of the deformed theory, we
can check this invariance on the dual field strengths of the remaining
fields\footnote{We leave aside the scalars for the moment.}, which is
tantamount to checking the invariance of the equations of motion of the
fundamental fields.

Using the duality relation $R_{a}=G_{a}$ the equations of motion of the
1-forms can be written in the form

\begin{equation}
\frac{\delta S}{\delta A^{I}}
=
-d
\left\{  
\mathcal{M}_{IJ}\star F^{J} 
-d_{iI}{}^{m}G^{i}B_{m} 
-\tfrac{1}{2} d_{iI}{}^{m}d^{i}{}_{J}{}^{n}F^{J}B_{mn}
\right\}
=0\, ,
\end{equation}

\noindent
and can be solved by identifying all the terms inside the brackets with
$d\tilde{A}_{I}$, where $\tilde{A}_{I}$ is a set of 5-forms. Taking into
account gauge invariance, the 6-form field strengths $\tilde{F}_{I}$ have the
following definition, duality relation and Bianchi identities:

\begin{eqnarray}
\tilde{F}_{I}
& \equiv &
d\tilde{A}_{I}+d_{iI}{}^{m}G^{i}B_{m} 
+\tfrac{1}{2} d_{iI}{}^{m}d^{i}{}_{J}{}^{n}F^{J}B_{mn}\, ,  
\\
& & \nonumber \\
\tilde{F}_{I}
& = &  
\mathcal{M}_{IJ}\star F^{J}\, ,
\\
& & \nonumber \\
d\tilde{F}_{I}
& = &
d_{iI}{}^{m}G^{i}H_{m}\, ,  
\end{eqnarray}

\noindent
and the equations of motion are of the 1-forms given by the Bianchi identities
of the dual 6-form field strengths up to duality relations:

\begin{equation}
\frac{\delta S}{\delta A^{I}}
=
-\left\{
d\tilde{F}_{I}-d_{iI}{}^{m}G^{i}H_{m}
\right\}\, .  
\end{equation}

Using the duality relation $R_{a}=G_{a}$ and following the same steps for the
2-forms , we find

\begin{eqnarray}
\tilde{H}^{m}
& = & 
d\tilde{B}^{m} +d^{i}{}_{I}{}^{m}F^{I}C_{i}\, ,
\\
& & \nonumber \\
\tilde{H}^{m}
& = & 
\mathcal{M}^{mn}\star H_{n}\, ,
\\
& & \nonumber \\
d\tilde{H}^{m}
& = & 
-d_{iI}{}^{m}G^{i}F^{I}\, ,
\end{eqnarray}

\noindent
and the equations of motion of the 2-forms are given by the Bianchi identities
of the dual 5-form field strengths up to duality relations:

\begin{equation}
\frac{\delta S}{\delta B_{m}}
=
-\left\{
d\tilde{H}^{m}+d_{iI}{}^{m}G^{i}F^{I}
\right\}\, .  
\end{equation}

This completes the first Abelian deformation. The second non-trivial
deformation of $G^{a}$ that one could consider is the addition of a CS 4-form
term $\sim d^{a}{}_{IJK}A^{I}F^{J}A^{K}$. The gauge transformation of this
term is not a total derivative and we cannot make $G^{a}$ gauge-invariant by
deforming the gauge transformation rule of $C^{a}$ only: we must also deform
that of $B_{m}$, which, in its turn, induces a deformation of $H_{m}$ by
addition of a CS 3-form term. Since the deformation of $H_{m}$ is essentially
unique, it is more convenient to start from this side and redefine

\begin{equation}
\label{eq:Abelian3-formfieldstrength}
H_{m}= dB_{m} - d_{mIJ}F^{I}A^{J}\, ,
\end{equation}

\noindent
where $d_{mIJ}=d_{mJI}$\footnote{The antisymmetric part is a total derivative
  that can be absorbed into a redefinition of $B_{m}$.}  which is invariant
under the gauge transformations

\begin{equation}
\label{eq:Abelian1-ad2-formgaugetransformations}
\delta_{\sigma}A^{I} = d\sigma^{I}\, ,
\hspace{1cm}  
\delta_{\sigma}B_{m} = d\sigma_{m} + d_{mIJ}F^{I}\sigma^{J}\, ,
\end{equation}

\noindent
and satisfies the Bianchi identities

\begin{equation}
dH_{m}
=
-d_{mIJ}F^{IJ}\, .  
\end{equation}

Under these gauge transformations and a generic $\delta_{\sigma}C^{a}$

\begin{equation}
\delta_{\sigma}G^{a}
=
d\left(\delta_{\sigma}C^{a} + d^{a}{}_{I}{}^{m}F^{I}\sigma_{m} \right) 
+d^{a}{}_{I}{}^{m}d_{mJK} F^{IJ}\sigma^{K}\, . 
\end{equation}

\noindent
Adding a CS 4-form term to $G^{a}$ 

\begin{equation}
G^{a} 
= 
dC^{a} +d^{a}{}_{I}{}^{m}F^{I}B_{m} 
-\alpha d^{a}{}_{I}{}^{m}d_{mJK} A^{I}F^{J}A^{K}\, .
\end{equation}

\noindent
we find

\begin{equation}
\begin{array}{rcl}
\delta_{\sigma}G^{a}
& = &
d
\left[
\delta_{\sigma}C^{a}+d^{a}{}_{I}{}^{m}F^{I}\sigma_{m} 
-\alpha d^{a}{}_{I}{}^{m}d_{mJK}(\sigma^{I}F^{J}A^{K} - A^{I}F^{J}\sigma^{K}) 
\right]
\\
& & \\
& &  
+d^{a}{}_{I}{}^{m}d_{mJK}
\left[\alpha \sigma^{I} F^{JK} +(1-\alpha) F^{IJ}\sigma^{K}\right]\, . 
\end{array}
\end{equation}

The last term can be made to vanish by simply requiring

\begin{equation}
\alpha d^{a}{}_{I}{}^{m}d_{mJK}  = (\alpha-1)d^{a}{}_{(J|}{}^{m}d_{m|K)I}\, .
\end{equation}

Symmetrizing both sides of this equation w.r.t.~$IJK$ we conclude that 

\begin{equation}
\label{eq:ddidentity0}
d^{a}{}_{(I|}{}^{m}d_{m|JK)}=0\, ,  
\end{equation}

\noindent 
and going back to the original (unsymmetrized) equation this implies that
$\alpha=1/3$.  We arrive to the field strength, gauge transformation and
Bianchi identities

\begin{eqnarray}
G^{a} 
& = & 
dC^{a} +d^{a}{}_{I}{}^{m}F^{I}B_{m} 
-\tfrac{1}{3} d^{a}{}_{I}{}^{m}d_{mJK} A^{I}F^{J}A^{K}\, ,
\\
& & \nonumber \\
\delta_{\sigma}C^{a}
& = &
d\sigma^{a}
-d^{a}{}_{I}{}^{m}F^{I}\sigma_{m} 
+\tfrac{1}{3} d^{a}{}_{I}{}^{m}d_{mJK}
(\sigma^{I}F^{J}A^{K} - A^{I}F^{J}\sigma^{K})\, ,  
\\
& & \nonumber \\
dG^{a} 
& = & 
d^{a}{}_{I}{}^{m}F^{I}H_{m}\, .
\end{eqnarray}

If these deformations are going to preserve formal symplectic invariance, we
expect that these results extend to the dual 3-forms and 4-forms field
strengths, that is:

\begin{eqnarray}
\label{eq:Abelian4-formfieldstrength}
G^{i} 
& = & 
dC^{i} +d^{i}{}_{I}{}^{m}F^{I}B_{m} 
-\tfrac{1}{3} d^{i}{}_{I}{}^{m}d_{mJK} A^{I}F^{J}A^{K}\, ,
\\
& & \nonumber \\
\delta_{\sigma}C^{i}
& = &
\label{eq:Abelian3-formgaugetransformartions}
d\sigma^{i}
-d^{i}{}_{I}{}^{m}F^{I}\sigma_{m} 
+\tfrac{1}{3} d^{i}{}_{I}{}^{m}d_{mJK}
(\sigma^{I}F^{J}A^{K} - A^{I}F^{J}\sigma^{K})\, ,  
\\
& & \nonumber \\
dG^{i} 
& = & 
d^{i}{}_{I}{}^{m}F^{I}H_{m}\, ,
\end{eqnarray}

\noindent
while the identity

\begin{equation}
\label{eq:ddidentity}
d^{i}{}_{(I|}{}^{m}d_{m|JK)}=0\, .
\end{equation}

This requires the introduction of new CS terms in the action. If we define the
CS terms in the 4-form field strengths by $\Delta G^{i}$ ($G^{i}=dC^{i}+\Delta
G^{i}$), then we expect the following terms to be present:

\begin{equation}
\label{eq:CS3}
S_{CS}
= 
-\int 
\{ 
dC^{a}\Delta G_{a} +\tfrac{1}{2}\Delta G^{a} \Delta G_{a}
\}\, .  
\end{equation}

\noindent
Instead of checking in detail the gauge-invariance of these terms, it is more
convenient to take the formal exterior derivative and check whether it is
entirely given in terms of the gauge-invariant field strengths found above. if
it is not, it should fail only by a total derivative which we can compensate
by adding the corresponding terms to the action.

We have found that one has to relate $d^{i}{}_{(I|}{}^{m}d_{i|J)}{}^{n}$ to
the tensor $d_{mIJ}$. The relation can be established by introducing a new
tensor $d^{mnp}= -d^{nmp}$ and is given by

\begin{equation}
\label{eq:relationdddd}
d^{i}{}_{(I|}{}^{m}d_{i|J)}{}^{n}
=
-2d^{mnp}d_{pIJ}\, .  
\end{equation}

\noindent
Observe that $d^{i}{}_{[I|}{}^{m}d_{i|J]}{}^{n}$ does not necessarily vanish.

Using the above relation we find a result of the expected form\footnote{We use
repeatedly the identity
\begin{equation}
2d^{i}{}_{I}{}^{m}d_{iJ}{}^{n}F^{I}A^{J}\Delta H_{n}
= 
-6 d^{mnp}\Delta H_{n} \Delta H_{p}
+d\{ \tfrac{1}{2}d^{i}{}_{I}{}^{m}d_{iJ}{}^{n} A^{IJ} \Delta H_{n} \}\, .  
\end{equation}
}

\begin{equation}
\begin{array}{rcl}
d
\{ 
dC^{a}\Delta G_{a} +\tfrac{1}{2}\Delta G^{a} \Delta G_{a}
\}
& = &
d_{aI}{}^{m}G^{a}F^{I}H_{m} -\tfrac{1}{3}d^{mnp}H_{mnp}
+
d 
\left\{
- \tfrac{1}{6}
d^{mnp}B_{m}
dB_{n}dB_{p}
\right.
\\
& & \\
& & 
\left.
+\tfrac{1}{2}d^{mnp}B_{m}H_{np} 
+\tfrac{1}{24}d^{i}{}_{I}{}^{m}d_{iJ}{}^{n} A^{IJ}\Delta H_{m}dB_{n}
\right\}\, ,
\end{array}
\end{equation}

\noindent
from which it follows that the gauge-invariant CS term in the action is given,
up to total derivatives, by

\begin{equation}
  \begin{array}{rcl}
S_{CS}
& = &
{\displaystyle \int }
\left\{
-dC^{a}\Delta G_{a} -\tfrac{1}{2}\Delta G^{a} \Delta G_{a}
-\tfrac{1}{6}d^{mnp}B_{m}dB_{n}dB_{p}
+\tfrac{1}{2}d^{mnp}B_{m}H_{np} 
\right.
\\
& & \\
& & 
\left.
+\tfrac{1}{24}d^{i}{}_{I}{}^{m}d_{iJ}{}^{n} A^{IJ}\Delta H_{m}dB_{n}
\right\}\, .  
\end{array}
\end{equation}

Observe that only the completely antisymmetric part of $d^{mnp}$ enters the
action, even though we have only assumed it to be antisymmetric in the first
two indices. We will henceforth assume that $d^{mnp}$ is completely
antisymmetric.

Now, as a final check of the consistency of our results, we can compute the
dual field strengths $\tilde{H}^{m}$ and $\tilde{F}_{I}$, which should be
formally symplectic invariant if the theory is, and their Bianchi identities,
which should be given entirely in terms of other field strengths if the theory
is indeed gauge invariant.

We find 

\begin{eqnarray}
\tilde{H}^{m}
& = & 
d\tilde{B}^{m} 
+d^{i}{}_{I}{}^{m} C_{i}F^{I}  
+d^{mnp}B_{n}(H_{p}+\Delta H_{p})
+\tfrac{1}{12}d^{i}{}_{I}{}^{m}d_{iJ}{}^{n} A^{IJ}\Delta H_{n}\, ,
\\
& & \nonumber \\
d\tilde{H}^{m}
& = & 
d^{i}{}_{I}{}^{m}G_{i}F^{I} + d^{mnp}H_{np}\, ,
\\
& & \nonumber \\
\tilde{F}_{I}
& = &
d\tilde{A}_{I} +2d_{mIJ}A^{J}(\tilde{H}_{m}-\tfrac{1}{2}\Delta\tilde{H}_{m})
-\left( 
d^{i}{}_{I}{}^{m}B_{m} -\tfrac{1}{3}d^{i}{}_{J}{}^{m}d_{mIK}A^{JK}
\right)
(G_{i}-\tfrac{1}{2}\Delta G_{i})
\nonumber
\\
& & \nonumber \\
& & 
-\tfrac{1}{3}
\left(
d^{i}{}_{I}{}^{m}d_{mJK} -d^{i}{}_{K}{}^{m}d_{mIJ}
\right) 
F^{J}A^{K}C_{i}
-d^{mnp}d_{mIJ}A^{J}B_{n}H_{p}
\nonumber 
\\
& & \nonumber \\
& & 
+\tfrac{1}{24}
\left(
d^{i}{}_{K}{}^{m}d_{iL}{}^{n}d_{mIJ}
+2d^{i}{}_{[I|}{}^{m}d_{i|K]}{}^{n}d_{mJL}
\right)
F^{J}A^{KL}B_{n}
+\tfrac{1}{24} d^{i}{}_{J}{}^{m} d_{iK}^{n} d_{mIL}A^{JKL}dB_{n}
\nonumber 
\\
& & \nonumber \\
& & 
-\tfrac{1}{180} d^{i}{}_{L}{}^{n}d_{iQ}{}^{m}d_{mIJ} d_{nPK} A^{JKLQ}F^{P}\, ,
\\
& & \nonumber \\
d\tilde{F}_{I}
& = &
2d_{mIJ}F^{J}\tilde{H}^{m} +d_{iI}{}^{m}G^{i}H_{m}\, .
\end{eqnarray}

\noindent
The duality relations are the same as in the undeformed case. 

As a further check of this construction, taking the exterior derivative of the
Bianchi identities of all the field strengths one finds consistent results
upon use of the properties of the deformation tensors
$d^{i}{}_{I}{}^{m},d_{mIJ},d^{mnp}$.

We will not compute the gauge transformations of the higher-rank form fields
since they will not be necessary in what follows.

\subsection{The 6-form potentials and their 7-form field strengths}
\label{sec-6-formpotentials}

On general grounds (see \cite{Bandos:2016smv} and references therein) the 6-form
potentials are expected to be the duals of the scalars. However, maintaining
the manifest invariances of the theory in the dualization procedure requires
the introduction of as many 6-forms $D_{A}$ as generators of global
transformations $\delta_{A}$ leaving the equations of motion (not just the
action) invariant. Hence, the index $A$ labels the adjoint representation of
the duality group. The 7-form field strengths $K_{A}$ are the Hodge duals of
the piece $j^{(\sigma)}_{A}(\phi)$ of the Noether--Gaillard--Zumino (NGZ)
conserved 1-form currents $j_{A}=j^{(\sigma)}_{A}(\phi)+\Delta j_{A}$
associated to those symmetries (or, better, dualities) \cite{Gaillard:1981rj}
which only depend on the scalar fields\footnote{This is the contribution of
  the $\sigma$-model to the Noether current. The symmetries of the equations
  of motion are necessarily symmetries of the $\sigma$-model, i.e.~isometries
  of the $\sigma$-model metric $\mathcal{G}_{xy}(\phi)$ generated by Killing
  vectors $k_{A}{}^{x}$. The indices $A,B,C$ label the symmetries of the
  theory and, therefore, run over the adjoint representation of the Lie
  algebra of that symmetry group G. The contribution of the $\sigma$-model to
  the NGZ 1-form is $j^{(\sigma)}_{A} = k_{A}{}^{x}\mathcal{G}_{xy}d\phi^{y}$.}

\begin{equation}
 K_{A}\equiv -\star j^{(\sigma)}_{A}\, , 
\end{equation}

\noindent
and their Bianchi identities follow from the conservation law for those
currents

\begin{equation}
d K_{A} =d \star j^{(\sigma)}_{A}= d\star (j^{NGZ}_{A}- \Delta j_{A})= 
-d\star \Delta j_{A}\, ,
\end{equation}

\noindent
where we have used the conservation of the NGZ current.

The simplest procedure to compute $\Delta j_{A}$ is to contract the equations
of motion of the scalars with the Killing vectors $k_{A}{}^{x}(\phi)$ of the
$\sigma$-model metric $\mathcal{G}_{xy}(\phi)$, which is given by 

\begin{equation}
  \begin{array}{rcl}
{\displaystyle\frac{\delta S}{\delta \phi^{x}}}
& = &
-d(\star \mathcal{G}_{xy}d\phi^{y}) +\tfrac{1}{2} \partial_{x}
\mathcal{G}_{yz}d\phi^{y}\wedge \star d\phi^{z}
\\
& & \\
& & 
+\tfrac{1}{2}\partial_{x}\mathcal{M}_{IJ}F^{I}\wedge \star F^{J}
+\tfrac{1}{2}\partial_{x}\mathcal{M}^{mn}H_{m}\wedge \star H_{n}
+G^{a}\partial_{x} G_{a}\, .
\end{array}
\end{equation}

\noindent
Using the Killing equation, we get

\begin{equation}
k_{A}{}^{x}\frac{\delta S}{\delta \phi^{x}}
 =
-d\star j_{A}^{(\sigma)}
+\tfrac{1}{2}k_{A}{}^{x}\partial_{x}\mathcal{M}_{IJ}F^{I}\wedge \star F^{J}
+\tfrac{1}{2}k_{A}{}^{x}\partial_{x}\mathcal{M}^{mn}H_{m}\wedge \star H_{n}
+G^{a}k_{A}{}^{x}\partial_{x} G_{a}\, .
\end{equation}

We must now use the fact that the isometry generated by $k_{A}$ will only be a
symmetry of the equations of motion if\footnote{The transformation rule for
  the period matrix is unconventional because our definition of the lower
  component of the symplectic vector of 4-form field strengths, $G_{a}=R_{a}$
  is unconventional (the sign is the oposite to the conventional one).}

\begin{equation}
  \begin{array}{rcl}
k_{A}{}^{x}\partial_{x}\mathcal{M}_{IJ} 
& = & 
-2T_{A}{}^{K}{}_{(I}\mathcal{M}_{J)K}\, ,
\\
& & \\
k_{A}{}^{x}\partial_{x}\mathcal{M}^{mn} 
& = & 
2T_{A}{}^{(m}{}_{p}\mathcal{M}^{n)p}\, ,
\\
& & \\
k_{A}{}^{x}\partial_{x} \mathcal{N}_{ab}  
& = &
-T_{A\, ab} 
-\mathcal{N}_{ac}T_{A}{}^{c}{}_{b}
+T_{A\, a}{}^{c}\mathcal{N}_{cb}
+\mathcal{N}_{ac}T_{A}{}^{cd}
\mathcal{N}_{db}\, ,
\end{array}
\end{equation}

\noindent
where the matrices $T_{A}{}^{I}{}_{J},T_{A}{}^{m}{}_{n}$ and 

\begin{equation}
\left( T_{A}{}^{i}{}_{j}   \right)
\equiv 
\left(
  \begin{array}{cc}
T_{A}{}^{a}{}_{b} & T_{A}{}^{ab} \\
T_{A\, ab} & T_{A\, a}{}^{b} \\
  \end{array}
\right)\, ,
\end{equation}

\noindent
are generators of the symmetry group G in the representation in which the
1-forms, 2-forms and 3-forms transform

\begin{equation}
[T_{A},T_{B}] = f_{AB}{}^{C}T_{C}\, ,
\hspace{1cm}
[k_{A},k_{B}] = -f_{AB}{}^{C}k_{C}\, .  
\end{equation}

\noindent
As we have discussed, this implies that the matrices $T_{A}{}^{i}{}_{j}$ are
generators of the symplectic group

\begin{equation}
T_{A}{}^{i}{}_{[j}\Omega_{k]i} = 0\, .  
\end{equation}

Upon use of the duality relations between field strengths, we find that 

\begin{equation}
-k_{A}{}^{x}\frac{\delta S}{\delta \phi^{x}}
 =
d\star j_{A}^{(\sigma)} +T_{A}{}^{J}{}_{I}F^{I}\tilde{F}_{J}
+T_{A}{}^{m}{}_{n}\tilde{H}^{n}H_{m}
-\tfrac{1}{2}T_{Aij}G^{ij}=0\, ,
\end{equation}

\noindent
on shell. The exterior derivative of the whole expression vanishes due to the
Bianchi identities of the field strengths and to the invariance of the
deformation tensors $d_{mIJ},d^{i}{}_{I}{}^{m}$ and $d^{mnp}$ under the
$\delta_{A}$ transformations:

\begin{equation}
\begin{array}{rcl}
\delta_{A}d_{mIJ} 
& = &
-T_{A}{}^{n}{}_{m} d_{nIJ}
-2T_{A}{}^{K}{}_{(I|} d_{n|J)K}=0\, ,
\\
& & \\
\delta_{A}d^{i}{}_{I}{}^{m}
& = &
T_{A}{}^{i}{}_{j} d^{j}{}_{I}{}^{m}
-T_{A}{}^{J}{}_{I} d^{i}{}_{J}{}^{m}
+T_{A}{}^{m}{}_{n} d^{i}{}_{I}{}^{n}=0\, ,
\\
& & \\
\delta_{A} d^{mnp}
& = &
3T_{A}{}^{[m|}_{q}d^{q|np]}=0\, .
\end{array}
\end{equation}

\noindent
This means that we can rewrite that equation locally as the conservation of
the NGZ current

\begin{equation}
d\star j_{A}^{NGZ} =0\, ,  
\hspace{1cm}
j_{A}^{NGZ} \equiv j^{(\sigma)}_{A}+\Delta j_{A}\, ,
\end{equation}

\noindent
where $\Delta j_{A}$ is a very long and complicated expression whose explicit
form will not be useful for us.  A local solution is provided by
$\star[j^{(\sigma)}_{A}+\Delta j_{A}] = -d D_{A}$ for the 6-form potential
$D_{A}$ and we get the definition of the 7-form field strength

\begin{equation}
\star j^{(\sigma)}_{A}= -dD_{A} +\star\Delta j_{A} \equiv K_{A}\, .
\end{equation}

\noindent
Its Bianchi identity is given by

\begin{equation}
dK_{A}
=
-d\star j^{(\sigma)}_{A}
=
T_{A}{}^{I}{}_{J}F^{J}\tilde{F}_{I}
+T_{A}{}^{m}{}_{n}\tilde{H}^{n}H_{m}
+\tfrac{1}{2}T_{Aij}G^{ij}\, .
\end{equation}

In the kind of theories that we are considering here there is no reason to
include potentials of rank higher than 6, unless we introduce a scalar
potential depending on new coupling constants: one can then introduce 7-form
potentials dual to those coupling constants. Since the introduction of these
parameters would be purely \textit{ad hoc}, we will postpone the study of this
duality to the next section in which we will be able to use in the definition
of the scalar potential the embedding tensor and the massive deformation
parameters, which have well-defined properties.

One can also generalize the theory by adding a scalar potential. This addition
is associated to the introduction of new deformation parameters.  In gauged
supergravity, which is the main case of interest, these deformation parameters
are the components of the embedding tensor and the scalar potential arises in
the gauging procedure, associated to the fermion shifts in the fermion's
supersymmetry transformations. Thus, it is natural to deal with the scalar
potential in the next section too.

The results obtained in this and the previous Section are summarized in
Appendix~\ref{app-genericmassless}.

\section{Non-Abelian and  massive deformations: the tensor hierarchy}
\label{sec-nonabelian}

The next step in the construction of the most general $d=8$ field theory is
the gauging of the global symmetries of the theory.  The most general
possibilities can be explored using the embedding tensor formalism\footnote{In
  this section we will follow Ref.~\cite{Ortin:2015hya}, where the essential
  references on the embedding tensor formalism can be found.  We will also use
  the same notation.} and in this section we are going to set it up for the
Abelian theories we have just found.\footnote{Observer that, in general, the
  theories that we are considering are just the bosonic sector of a theory
  that also contains fermions and whose symmetry group may include symmetries
  that only act on them. The total symmetry group would, then, be larger and
  the embedding tensor should take this fact into account.} For the sake of
convenience we are going to reproduce some of the formulae obtained above.

The starting point is the assumption that the equations of motion of the
theory are invariant under a global symmetry group with infinitesimal
generators $\{T_{A}\}$ satisfying the algebra

\begin{equation}
[T_{A},T_{B}] = f_{AB}{}^{C}T_{C}\, .
\end{equation}

\noindent
The group acts linearly on all the forms of rank $\geq 1$, including the
3-forms if the electric and magnetic 3-forms $C^{a}$ and $C_{a}$ are combined
into a single symplectic vector of 3-forms 
$
(C^{i})
=
\left(
\begin{smallmatrix}
  C^{a}\\ C_{a} \\
\end{smallmatrix}
\right) $ as explained above and codify electric-magnetic transformations
involving the scalars. The matrices that represent the generators are denoted
by $\{T_{A}{}^{I}{}_{J}\},\{T_{A}{}^{m}{}_{n}\},\{T_{A}{}^{i}{}_{j}\}$ and the
adjoint generators are $T_{A}{}^{B}{}_{C}= f_{AC}{}^{B}$. The matrices
$T_{A}{}^{i}{}_{j}$ are generators of the symplectic group

\begin{equation}
T_{A}{}^{i}{}_{[j}\Omega_{k]i} = 0\, ,
\hspace{1cm}
( \Omega_{ij}) 
= 
(\Omega^{ij}) 
\equiv  
  \left(
  \begin{array}{cc}
   0 & \mathbbm{1} \\
-\mathbbm{1} & 0 \\ 
  \end{array}
\right)\, .  
\end{equation}

\noindent
We have

\begin{equation}
\begin{array}{rclrclrcl}
\delta_{\alpha} A^{I} 
& = & 
\alpha^{A}T_{A}{}^{I}{}_{J}A^{J}\, ,
\hspace{1cm}
&
\delta_{\alpha} B_{m} 
& = &
-\alpha^{A}T_{A}{}^{n}{}_{m}B_{n}\, ,
\hspace{1cm}
&
\delta_{\alpha} C^{i} 
& = &
\alpha^{A}T_{A}{}^{i}{}_{j}C^{j}\, ,
\\
& & & & & & & & \\
\delta_{\alpha} \tilde{A}_{I} 
& = & 
-\alpha^{A}T_{A}{}^{J}{}_{I}\tilde{A}_{J}\, ,
\hspace{1cm}
&
\delta_{\alpha} \tilde{B}^{m} 
& = &
\alpha^{A}T_{A}{}^{m}{}_{n}\tilde{B}^{n}\, ,
\hspace{1cm}
& & & \\
\end{array}
\end{equation}

\noindent
(the dual potentials transform in the dual covariant-contravariant representation).

The kinetic matrices
$\mathcal{M}_{IJ},\mathcal{M}^{mn},\mathcal{W}_{ij}(\mathcal{N})$ also
transform linearly: if $\delta_{\alpha}\equiv \alpha^{A}\delta_{A}$

\begin{equation}
\delta_{A}\mathcal{M}_{IJ} 
= 
-2T_{A}{}^{K}{}_{(I}\mathcal{M}_{J)K}\, ,
\hspace{.5cm}
\delta_{A}\mathcal{M}^{mn} 
= 
2T_{A}{}^{(m}{}_{p}\mathcal{M}^{n)p}\, ,
\hspace{.5cm}
\delta_{A}\mathcal{W}_{ij} 
= 
-2T_{A}{}^{k}{}_{(i}\mathcal{W}_{j)k}\, ,
\end{equation}

\noindent
but the period matrix undergoes fractional-linear transformations which,
infinitesimally, take the form

\begin{equation}
\delta_{A} \mathcal{N}_{ab}  
=
-T_{A\, ab} 
-\mathcal{N}_{ac}T_{A}{}^{c}{}_{b}
+T_{A\, a}{}^{c}\mathcal{N}_{cb}
+\mathcal{N}_{ac}T_{A}{}^{cd}
\mathcal{N}_{db}\, ,
\hspace{1cm}
\left( T_{A}{}^{i}{}_{j}   \right)
=
\left(
  \begin{array}{cc}
T_{A}{}^{a}{}_{b} & T_{A}{}^{ab} \\
T_{A\, ab} & T_{A\, a}{}^{b} \\
  \end{array}
\right)\, .
\end{equation}

The $k$-form field strengths will transform in the same representation as the
corresponding $(k-1)$-form potential, but only if the $d$-tensors
$d_{mIJ},d^{i}{}_{I}{}^{m}$ and $d^{mnp}$ are invariant under the global
symmetry group, \textit{i.e.}~if they must satisfy

\begin{equation}
\begin{array}{rcl}
\delta_{A}d_{mIJ} 
& = &
-T_{A}{}^{n}{}_{m} d_{nIJ}
-2T_{A}{}^{K}{}_{(I|} d_{n|J)K}=0\, ,
\\
& & \\
\delta_{A}d^{i}{}_{I}{}^{m}
& = &
T_{A}{}^{i}{}_{j} d^{j}{}_{I}{}^{m}
-T_{A}{}^{J}{}_{I} d^{i}{}_{J}{}^{m}
+T_{A}{}^{m}{}_{n} d^{i}{}_{I}{}^{n}=0\, ,
\\
& & \\
\delta_{A} d^{mnp}
& = &
3T_{A}{}^{[m|}_{q}d^{q|np]}=0\, .
\end{array}
\end{equation}

The theories we have constructed are invariant under Abelian gauge
transformations with $0$-, $1$- and $2$-form parameters
$\sigma^{I},\sigma_{m},\sigma^{i}$:

\begin{equation}
\delta_{\sigma}A^{I} \sim d\sigma^{I}\, ,
\hspace{1cm}  
\delta_{\sigma}B_{m} \sim d\sigma_{m}\, ,
\hspace{1cm}  
\delta_{\sigma}C^{i} \sim d\sigma^{i}\, .
\end{equation}

In order to gauge the global symmetries, we promote the global parameters
$\alpha^{A}$ to local ones $\alpha^{A}(x)$ and we identify them with some
combinations of the gauge parameters of the 1-forms $\sigma^{I}$ via the
embedding tensor $\vartheta_{I}{}^{A}$ as follows:

\begin{equation}
\alpha^{A}\equiv \sigma^{I}\vartheta_{I}{}^{A}\, .  
\end{equation}

\noindent
Using this redefinition in the transformation of the kinetic matrices
$\mathcal{M}_{IJ},\mathcal{M}^{mn},\mathcal{W}_{ij}$ one immediately finds their gauge
transformations:

\begin{equation}
\delta_{\sigma}\mathcal{M}_{IJ} 
= 
-2\sigma^{L} X_{L}{}^{K}{}_{(I}\mathcal{M}_{J)K}\, ,
\hspace{.5cm}
\delta_{\sigma}\mathcal{M}^{mn} 
= 
2\sigma^{I}T_{I}{}^{(m}{}_{p}\mathcal{M}^{n)p}\, ,
\hspace{.5cm}
\delta_{\sigma}\mathcal{W}_{ij} 
= 
-2\sigma^{I}X_{I}{}^{k}{}_{(i}\mathcal{W}_{j)k}\, ,
\end{equation}

\noindent
where we have defined the matrices

\begin{equation}
X_{I}{}^{J}{}_{K} \equiv \vartheta_{I}{}^{A} T_{A}{}^{J}{}_{K}\, ,
\hspace{.5cm} 
X_{I}{}^{m}{}_{n} \equiv \vartheta_{I}{}^{A} T_{A}{}^{m}{}_{n}\, . 
\hspace{.5cm} 
X_{I}{}^{i}{}_{j} \equiv \vartheta_{I}{}^{A} T_{A}{}^{i}{}_{j}\, . 
\end{equation}

The gauge fields for these symmetries are given by

\begin{equation}
A^{A}\equiv A^{I}\vartheta_{I}{}^{A}\, .  
\end{equation}

\noindent
With them we can construct gauge-covariant derivatives, which we will then use
to derive Bianchi identities.

Is is convenient to start by constructing the covariant derivatives of the
kinetic matrices
$\mathcal{M}_{IJ},\mathcal{M}^{mn},\mathcal{W}_{ij}(\mathcal{N})$ which
transform linearly.  According to the general rule, the covariant derivative
of a field $\Phi$ transforming as $\delta_{A}\Phi$ is given by

\begin{equation}
\mathcal{D}\Phi \equiv d\Phi -A^{A}\delta_{A}\Phi\, .
\end{equation}

Then, with the above definition of  gauge fields 

\begin{eqnarray}
\mathcal{D}\mathcal{M}^{mn}
& = & 
d\mathcal{M}^{mn}
-2A^{I}X_{I}{}^{(m}{}_{p}\mathcal{M}^{n)p}\, ,
\\
& & \nonumber \\
\mathcal{D}\mathcal{M}_{IJ}
& = & 
d\mathcal{M}_{IJ}
+2A^{L}X_{L}{}^{K}{}_{(I}\mathcal{M}_{J)K}\, ,
\\
& & \nonumber \\
\mathcal{D}\mathcal{W}_{ij}
& = & 
d\mathcal{W}_{ij}
+2A^{I}X_{I}{}^{k}{}_{(i}\mathcal{W}_{j)k}\, .
\end{eqnarray}

These derivatives transform covariantly under gauge transformations
$\delta_{\sigma}= \sigma^{I}\vartheta_{I}{}^{A}\delta_{A}$ provided that the
embedding tensor is gauge-invariant

\begin{equation}
\label{eq:deltastheta}
\delta_{\sigma}\vartheta_{I}{}^{A}
=
0\, ,  
\end{equation}

\noindent
and provided that the 1-forms transform as 

\begin{equation}
\delta_{\sigma}A^{I} = \mathcal{D} \sigma^{I} +\Delta_{\sigma} A^{I}\, ,
\,\,\,\,\,\,
\mbox{where}
\,\,\,\,\,\,
\left\{
\begin{array}{rcl}
\Delta_{\sigma} A^{I} \vartheta_{I}{}^{A} 
& = & 0\, , 
\\
& & \\
\mathcal{D} \sigma^{I} 
& = & 
d \sigma^{I} 
- A^{J} X_{J}{}^{I}{}_{K}\sigma^{K}\, ,
\end{array}
\right. 
\end{equation}

The condition Eq.~(\ref{eq:deltastheta}) leads to the so-called
\textit{quadratic constraint}

\begin{equation}
\label{eq:constraint1}
 \vartheta_{J}{}^{B}
\left[
T_{B}{}^{K}{}_{I}\vartheta_{K}{}^{A} -f_{BC}{}^{A}\vartheta_{I}{}^{C}
\right]
=
0\, .
\end{equation}

To determine $\Delta_{\sigma} A^{I}$ we have to construct the gauge-covariant 2-form
field strengths $F^{I}$. 

\subsection{2-form field strengths}

The simplest way to find the 2-form field strengths $F^{I}$ is through the
Ricci identities. A straightforward calculation using the quadratic constraint
Eq.~(\ref{eq:constraint1}) leads to

\begin{equation}
\label{eq:Ricci}
\mathcal{D}\mathcal{D}\mathcal{M}_{mn} =
-F^{I}\vartheta_{I}{}^{A}\delta_{A}\mathcal{M}_{mn}\, ,  
\end{equation}

\noindent
and analogous equations for $\mathcal{M}_{IJ}$ and
$\mathcal{W}_{ij}(\mathcal{N})$, with

\begin{equation}
F^{I}
=
dA^{I} -\tfrac{1}{2}X_{J}{}^{I}{}_{K}A^{JK} +\Delta F^{I}\, ,
\,\,\,\,\,\,
\mbox{where}
\,\,\,\,\,\,
\Delta F^{I}\vartheta_{I}{}^{A}=0\, .
\end{equation}

Under gauge transformations,

\begin{equation}
\delta_{\sigma}F^{I}
= 
\sigma^{J}X_{J}{}^{I}{}_{K}F^{K}\, ,
\end{equation}

\noindent
provided that 

\begin{equation}
\label{eq:condition}
\delta_{\sigma}\Delta F^{I} 
=  
-\mathcal{D}\Delta_{\sigma} A^{I}
+2X_{(J}{}^{I}{}_{K)}\left(F^{J} \sigma^{K}
  -\tfrac{1}{2}A^{J}\delta_{\sigma} A^{K} \right)\, .
\end{equation}

Given the field content of the theory, the natural candidate to $\Delta F^{I}$
and $\Delta_{\sigma} A^{I}$ are

\begin{equation}
\Delta F^{I} = Z^{I\,  m}B_{m}\, ,  
\hspace{1cm}
\Delta_{\sigma} A^{I} = -Z^{I\,  m}\sigma_{m}\, ,  
\end{equation}

\noindent
where the new tensor $Z^{I\, m}$ is gauge-invariant and orthogonal to the
embedding tensor:

\begin{eqnarray}
\delta_{\sigma} Z^{I\, m}
& = & 
0\, ,
\\
& & \nonumber \\
\label{eq:constraint2}
Z^{I\, m}\vartheta_{I}{}^{A}
& = & 
0\, .
\end{eqnarray}

\noindent
Then, the consistency of Eq.~(\ref{eq:condition}) with the above choice requires

\begin{equation}
\label{eq:constraint3}
\vartheta_{(J|}{}^{A}T_{A}{}^{I}{}_{|K)} = Z^{I\, m}d_{mJK}\, ,
\end{equation}

\noindent
for some tensor $d_{mJK}= d_{mKJ}$ which will turn out to coincide with the
tensor we introduced as an Abelian deformation in the previous
sections. Since we have assumed $\vartheta_{I}{}^{A}$ and $d_{mJK}$ to be
gauge-invariant, $Z^{I\, m}$ is automatically gauge-invariant and we have one
constraint less.

We conclude that\footnote{On general grounds, we expect a term of the form
  $-\sigma^{I}X_{I}{}^{n}{}_{m}B_{n}$ in the gauge transformation rule of
  $B_{m}$. This term is indeed present, but in a disguised form.}

\begin{eqnarray}
F^{I}
& = &
dA^{I} -\tfrac{1}{2}X_{J}{}^{I}{}_{K}A^{JK} +Z^{I\,  m}B_{m}\, ,
\\
& & \nonumber \\
\delta_{\sigma}F^{I}
& = &
\sigma^{J}X_{J}{}^{I}{}_{K}F^{K}\, ,
\\
& & \nonumber \\
\label{eq:nonAbelian1-formgaugetransformation}
\delta_{\sigma}A^{I} 
& = &
\mathcal{D} \sigma^{I} -Z^{I\,  m}\sigma_{m}\, ,
\\
& & \nonumber \\
\label{eq:nonAbelian2-formgaugetransformation}
\delta_{\sigma}B_{m} 
& = &  
\mathcal{D}\sigma_{m}
+2d_{mJK}\left(F^{J} \sigma^{K}
  -\tfrac{1}{2}A^{J}\delta_{\sigma}A^{K} \right)
+\Delta_{\sigma} B_{m}\, ,
\,\,\,
\mbox{with}
\,\,\,
Z^{I\,  m}\Delta_{\sigma} B_{m} =0\, .\hspace{1cm}~
\end{eqnarray}

In the ungauged limit $\vartheta_{I}{}^{A}=Z^{I\, m}=0$ we get the Abelian
gauge transformations of the 2-form
Eq.~(\ref{eq:Abelian1-ad2-formgaugetransformations}) if we identify the above
1-form $\sigma_{m}$ (relabeled $\sigma_{\mathrm{g}\, m}$)  with
$\sigma_{m}-d_{mIJ}A^{I}\sigma^{J}$, 

\begin{equation}
\label{eq:sigmagmversussigmam}
\sigma_{\mathrm{g}\, m}=\sigma_{m}-d_{mIJ}A^{I}\sigma^{J}\, ,
\end{equation}

\noindent
confirming the identification of the $d$-tensor. Using that variable makes the
non-Abelian gauge transformations more complicated and, therefore, we will
stick to the above $\sigma_{m}$.

\subsection{3-form field strengths}

Again, the shortest way to find $\Delta B_{m}$ and the gauge-covariant 3-form
field strength $H_{m}$ is through the Bianchi identities. Taking the covariant
derivative of the 2-form field strength, and using the generalized Jacobi
identity

\begin{equation}
X_{[I|}{}^{K}{}_{M}X_{|J}{}^{M}{}_{L]}
=
\tfrac{2}{3} Z^{K\, m}d_{mM[I}X_{J}{}^{M}{}_{L]}\, ,   
\end{equation}

\noindent
we get

\begin{equation}
\label{eq:Bianchigauged2}
\mathcal{D} F^{I} = Z^{I\, m}H_{m}\, ,   
\end{equation}

\noindent
where 

\begin{equation}
\label{eq:Hmgauged1}
H_{m}
=
\mathcal{D}B_{m} -d_{mIJ}dA^{I}A^{J} +\tfrac{1}{3}d_{mMI}X_{J}{}^{M}{}_{K}A^{IJK}
+\Delta H_{m}\, ,
\,\,\,\,\,\,
\mbox{with}
\,\,\,\,\,\,
Z^{I\,  m}\Delta H_{m} =0\, .
\end{equation}

In the ungauged limit $\vartheta_{I}{}^{A}=Z^{I\, m}=0$ we recover the Abelian
3-form field strength in Eq.~(\ref{eq:Abelian3-formfieldstrength}). On the
other hand, by construction, the above field strength is gauge-covariant up to
terms annihilated by $Z^{I\, m}$ under
Eqs.~(\ref{eq:nonAbelian1-formgaugetransformation}) and
(\ref{eq:nonAbelian2-formgaugetransformation}). To show this explicitly, we
will need further identities between the tensors of the theory that are more
easily discovered by computing first the 4-form field strengths.

\subsection{4-form field strengths}

From this moment, following Ref.~\cite{Hartong:2009vc}, we will determine the
general form of the field strengths using the Bianchi identities and their
consistency relations. This procedure yields gauge-covariant field strengths
and one can later find explicitly the gauge transformations of the fields that
produce that result.

Thus, we take the covariant derivative of both sides of
Eq.~(\ref{eq:Bianchigauged2}), use the Ricci identity Eq.~(\ref{eq:Ricci}) for
the l.h.s.~and the explicit form of $H_{m}$ in Eq.~(\ref{eq:Hmgauged1}) for
the r.h.s., and we find the Bianchi identity for $H_{m}$ to be

\begin{equation}
\mathcal{D}H_{m}  
=
-d_{mIJ}F^{IJ} +\Delta \mathcal{D}H_{m}\, ,
\,\,\,\,\,\,
\mbox{where}
\,\,\,\,\,\,
Z^{I\,  m}\Delta \mathcal{D}H_{m} =0\, .
\end{equation}

\noindent
$\Delta \mathcal{D}H_{m}$ has to be gauge-invariant and scalar-independent and
the only possibility is a 4-form combination of field strengths. $F^{I}\wedge
F^{J}$ has already been used and we must use $G^{i}$, whose explicit form will
be determined by consistency. We need to introduce a new gauge-invariant
tensor $Z_{im}$ orthogonal to $Z^{I\, m}$ 

\begin{equation}
\label{eq:constraint4}
Z^{I\, m}Z_{jm}
=  
0\, ,
\end{equation}

\noindent
and, then, we arrive to the Bianchi identity

\begin{equation}
\label{eq:Bianchigauged3}
\mathcal{D}H_{m}  
=
-d_{mIJ}F^{IJ} +Z_{im}G^{i}\, .
\end{equation}

A direct calculation of $\mathcal{D}H_{m}$ using the explicit expression of
$H_{m}$ in Eq.~(\ref{eq:Hmgauged1}) with $\Delta H_{m}= Z_{im}C^{i}$ can only
give a consistent result if we introduce a tensor $d^{i}{}_{I}{}^{m}$ such
that

\begin{equation}
\label{eq:constraint5}
X_{I}{}^{m}{}_{n} +2d_{nIJ}Z^{J\, m} = Z_{in}d^{i}{}_{I}{}^{m}\, .
\end{equation}

\noindent
The tensor $d^{i}{}_{I}{}^{m}$ coincides with the one we introduced as an
Abelian deformation. Also, observe that this relation makes the condition of
gauge invariance of $Z_{in}$ redundant.

We get

\begin{equation}
\label{eq:Gigauged1}
G^{i}
=
\mathcal{D}C^{i} 
+d^{i}{}_{I}{}^{n}
\left[
F^{I}B_{n} -\tfrac{1}{2}Z^{I\, p}B_{n}B_{p}
-\tfrac{1}{3}d_{nJK}dA^{J}A^{IK}
+\tfrac{1}{12}d_{nMJ}X_{K}{}^{M}{}_{L} A^{IJKL}
\right]
+\Delta G^{i}\, ,
\end{equation}

\noindent
with
 
\begin{equation}
Z_{im}\Delta G^{i} 
=
0\, .
\end{equation}

These 4-form field strengths reduce exactly to the Abelian ones in
Eq.~(\ref{eq:Abelian4-formfieldstrength}).

Now we are ready to check explicitly using the identity/constraint
Eq.~(\ref{eq:constraint5}) that $H_{m}$ in Eq.~(\ref{eq:Hmgauged1}) with
$\Delta H_{m} = Z_{im} C^{i}$ is gauge covariant up to terms proportional to
$Z_{im}$, which are automatically annihilated by $Z^{I\, m}$. We find that

\begin{equation}
  \begin{array}{rcl}
\delta_{\sigma} C^{i}
& = &
\mathcal{D}\sigma^{i}
-d^{i}{}_{I}{}^{n}
\left[
\sigma^{I}H_{n}
+F^{I}\sigma_{n}
+\delta_{\sigma}A^{I} B_{n}
-\tfrac{1}{3} d_{nJK}\delta_{\sigma}A^{J}A^{IK} 
\right]
+\Delta_{\sigma} C^{i}\, ,
\\
& & \\
& & 
\mbox{with}
\,\,\,\,\,
Z_{im}\Delta_{\sigma} C^{i} =0\, ,  
\\
& & \\
\Delta_{\sigma} B_{m}
& = &
-Z_{im}\sigma^{i}\, .
\end{array}
\end{equation}

These gauge transformations reduce to the Abelian ones
Eq.~(\ref{eq:Abelian3-formgaugetransformartions}) upon use of the property of
the $d$-tensors Eq.~(\ref{eq:ddidentity}) and the identifications
Eq.~(\ref{eq:sigmagmversussigmam}) and

\begin{equation}
\label{eq:sigmagiversussigmai}
\sigma_{\mathrm{g}}^{i}
=
\sigma^{i} +d^{i}{}_{I}{}^{n}(B_{n}\sigma^{I}-\tfrac{1}{3} d_{nJK}A^{IJ}\sigma^{K})\, .  
\end{equation}


\subsection{5-form field strengths}

Taking, once again, the covariant derivatives of both sides of the Bianchi
identity for $H_{m}$, Eq.~(\ref{eq:Bianchigauged3}), and using the Bianchi
identity for $F^{I}$, Eq.~(\ref{eq:Bianchigauged2}) and the newly introduced
tensor $d^{i}{}_{I}{}^{m}$, we find that 

\begin{equation}
\label{eq:Bianchigauged4}
\mathcal{D}G^{i} = d^{i}{}_{I}{}^{m} F^{I}H_{m} -Z^{i}{}_{m}\tilde{H}^{m}\, ,  
\end{equation}

\noindent
where $Z^{i}{}_{m}$ is a new gauge-invariant tensor orthogonal to $Z_{im}$ 

\begin{equation}
\label{eq:constraint6}
Z^{i}{}_{m}Z_{in}=0\, ,  
\end{equation}

\noindent
and where the sign of that term has been chosen so as to get the same signs as
in the ungauged case. In principle these two tensors could be completely
unrelated (except for the constraints). However, since, in the physical
theory, $G^{i}$ is self-dual and $\tilde{H}^{m}$ is the electric-magnetic dual
of $H_{m}$, it is natural to expect that the same tensors appear in both field
strengths. Thus, we are going to assume that $Z^{i}{}_{m}$ has been obtained
from $Z_{jm}$ by raising the index with the symplectic metric tensor
$\Omega^{ij}$, that is

\begin{equation}
Z^{i}{}_{m} \equiv \Omega^{ji}Z_{jm}\, .  
\end{equation}

\noindent
Then, there is no new constraint associated to its gauge invariance and, we
just have the constraint Eq.~(\ref{eq:constraint6}) analogous to a constraint
satisfied by the embedding tensor in 4-dimensional field theories.

\subsection{6-form field strengths}

Taking the covariant derivative of both sides of the Bianchi identity for
$G^{i}$, Eq.~(\ref{eq:Bianchigauged4}) and using the Bianchi identities for
the field strengths of lower rank, we find that we need to introduce three new
tensors $d_{iI}{}^{m}, d^{mnp}, d^{m}{}_{IJK}$ and demand that 

\begin{eqnarray}
\label{eq:constraint11}
d^{i}{}_{I}{}^{m}Z_{jm} +X_{I}{}^{i}{}_{j}
& = & 
-Z^{i}{}_{m}d_{jI}{}^{m}\, ,
\\
& & \nonumber \\
\label{eq:constraint8}
d^{i}{}_{I}{}^{[m|}Z^{I\, |n]}
& = & 
-Z^{i}{}_{p}d^{pmn}\, ,  
\\
& & \nonumber \\
\label{eq:constraint9}
d^{i}{}_{(I|}{}^{m}d_{m|JK)}
& = & 
-Z^{i}{}_{p} d^{p}{}_{IJK}\, .
\end{eqnarray}

\noindent
Lowering the $i$ indices in the first equation with $\epsilon_{ik}$ and taking
into account that $X_{I[kj]}=0$, we conclude that it is natural to identify

\begin{equation}
d_{iI}{}^{m}=\Omega_{ij}d^{j}{}_{I}{}^{m}\, ,
\end{equation}

\noindent
and rewrite the constraint as

\begin{equation}
\label{eq:constraint10}
X_{Iij} = -2 Z_{(i|m}d_{|j)I}{}^{m}\, .  
\end{equation}

Using these constraints and the same reasoning as in the previous cases we
find the next Bianchi identity and we can also solve it\footnote{Actually, it
  is easier to find $\tilde{H}^{M}$ from the previous Bianchi identity
  Eq.~(\ref{eq:Bianchigauged4}) taking the covariant derivative of the 4-form
  field strengths $G^{i}$ in Eq.~(\ref{eq:Gigauged1}) with $\Delta
  G^{i}=-Z^{i}{}_{m}\tilde{B}^{m}$.}

\begin{eqnarray}
\label{eq:Bianchigauged5}
\mathcal{D}\tilde{H}^{m}
& = &
-d_{iI}{}^{m}G^{i}F^{I} +d^{mnp}H_{np}  
+d^{m}{}_{IJK}F^{IJK} 
+Z^{I\, m}\tilde{F}_{I}\, ,
\\
& & \nonumber \\
\tilde{H}^{m}
& = &
\mathcal{D}\tilde{B}^{m} -d_{iI}{}^{m}F^{I}C^{i}
+2d^{mnp}B_{n}\left(H_{p}-Z_{ip}C^{i}-\tfrac{1}{2}\mathcal{D}B_{p}\right)
\nonumber \\
& & \nonumber \\
& & 
+d^{m}{}_{IJK}dA^{I}dA^{J}A^{K}
\nonumber \\
& & \nonumber \\
& & 
+
\left(
\tfrac{1}{12}d_{iJ}{}^{m}d^{i}{}_{K}{}^{n}d_{nIL} 
-\tfrac{3}{4} d^{m}{}_{IJM} X_{K}{}^{M}{}_{L}
\right)
dA^{I} A^{JKL}
\nonumber \\
& & \nonumber \\
& & 
+
\left(
\tfrac{3}{20} d^{m}{}_{NPM}X_{I}{}^{N}{}_{J}
-\tfrac{1}{60} d_{iM}{}^{m}d^{i}_{I}{}^{n}d_{nPJ}
\right) 
X_{K}{}^{P}{}_{L} A^{IJKLM}
\nonumber \\
& & \nonumber \\
& & 
+Z^{Im}\tilde{A}_{I}\, ,
\end{eqnarray}

\subsection{7-form field strengths}

Provided that we impose the additional constraint\footnote{This constraint
  reduces to Eq.~(\ref{eq:relationdddd}) in the ungauged, massless limit.}

\begin{equation}
\label{eq:constraint12}
d^{i}{}_{(I}{}^{m}d_{iJ)}{}^{n} +2 d^{mnp}d_{pIJ} +3d^{m}{}_{IJK} Z^{K\, n}
 =  
+3d^{n}{}_{IJK} Z^{K\, m}\, ,
\end{equation}

\noindent 
the covariant derivative of the Bianchi identity Eq.~(\ref{eq:Bianchigauged5})
leads to the Bianchi identity for the 6-form field strengths

\begin{eqnarray}
\label{eq:Bianchigauged6}
\mathcal{D}\tilde{F}_{I}
& = & 
2d_{mIJ} F^{J}\tilde{H}^{m} +d_{iI}{}^{m}G^{i}H_{m} 
-3 d^{m}{}_{IJK}F^{JK}H_{m} -\vartheta_{I}{}^{A}K_{A}\, ,
\\
& & \nonumber \\
\tilde{F}_{I}
& = &
\mathcal{D}\tilde{A}_{I} 
+2d_{mIJ}\tilde{B}^{m}F^{J}
+d_{iI}{}^{m}C^{i}(H_{m}-\tfrac{1}{2}Z_{jm}C^{j})
\nonumber \\
& & \nonumber \\
& & 
-3d^{m}{}_{IJK}B_{m}(F^{J}-\tfrac{1}{2}Z^{Jn}B_{n})(F^{K}-\tfrac{1}{2}Z^{Kp}B_{p})
-\tfrac{1}{4}d^{m}{}_{IJK}Z^{Jn}Z^{Kp}B_{mnp}
\nonumber \\
& & \nonumber \\
& & 
+\tfrac{1}{2}d_{iI}{}^{m}d^{i}{}_{J}{}^{n}(F^{J}-\tfrac{2}{3}Z^{Jp}B_{p})B_{mn}
-d_{iI}{}^{m}B_{m}\Box G^{i} +\cdots
\end{eqnarray}

\noindent
where we are denoting by $\Box G^{i}$ the part of $G^{i}$ that only contains
1-forms $A^{I}$ and their derivatives $dA^{I}$.

\subsection{8-form field strengths}
\label{sec-8-forms}

Taking the covariant derivative of Eq.~(\ref{eq:Bianchigauged6}) and using
several of the constraints imposed above, we find that 

\begin{equation}
\vartheta_{I}{}^{A}\mathcal{D}K_{A}
= 
X_{I}{}^{K}{}_{J}F^{J}\tilde{F}_{K}  
+X_{I}{}^{m}{}_{n}\tilde{H}^{n}H_{m}
+\tfrac{1}{2}X_{Iij}G^{ij}
+5d_{m(IJ}d^{m}{}_{KLM)}F^{JKLM}\, .
\end{equation}

According to the general arguments in Ref.~\cite{Bandos:2016smv} the last term must
vanish. It cannot arise in the Bianchi identity of the dual
Noether-Gaillard-Zumino current associated to the global symmetries of the
theory. Thus, we impose

\begin{equation}
\label{eq:constraint13}
d_{m(IJ}d^{m}{}_{KLM)} = 0\, ,  
\end{equation}

\noindent
and, from the definition of the $X$ tensors, we get

\begin{equation}
\mathcal{D}K_{A}
= 
T_{A}{}^{K}{}_{J}F^{J}\tilde{F}_{K}  
+T_{A}{}^{m}{}_{n}\tilde{H}^{n}H_{m}
-\tfrac{1}{2}T_{A\, ij}G^{ij}
+Y_{A}{}^{\sharp}L_{\sharp}\, ,
\end{equation}

\noindent
where $Y_{A}{}^{\sharp}$ is a tensor orthogonal to the embedding tensor

\begin{equation}
\label{eq:constraint14}
\vartheta_{I}{}^{A}Y_{A}{}^{\sharp}  =0\, ,
\end{equation}

\noindent
and where the index $\sharp$ runs over all the deformation tensors introduced
so far, that we are going to denote collectively by $c^{\sharp}$.  As argued
in Ref.~\cite{Hartong:2009vc}, the natural candidates for the
$Y_{A}{}^{\sharp}$ tensors are the variations of the deformation tensors
$c^{\sharp}$ under the global symmetries of the theory

\begin{equation}
Y_{A}{}^{\sharp} = \delta_{A}c^{\sharp}\, ,  
\end{equation}

\noindent
where $A$ runs over the whole Lie algebra of the global symmetry group,
because all the deformation tensors are required to be gauge invariant

\begin{equation}
\vartheta_{I}{}^{A}\delta_{A} c^{\sharp} = 
\vartheta_{I}{}^{A}Y_{A}{}^{\sharp}
\equiv
\mathcal{Q}_{I}{}^{\sharp}= 0\, ,  
\end{equation}

\noindent
where we have defined the constraints $\mathcal{Q}_{A}{}^{\sharp}$.

At this point there are two possibilities:

\begin{enumerate}
\item We can consider that all the independent tensors\footnote{The tensors
    $d^{mnp},d^{m}{}_{IJK}$ are related to these and their gauge invariance is
    not an independent condition.}
  $\{\vartheta_{I}{}^{A},Z^{Im},Z_{im},-d_{mIJ},d^{i}{}_{I}{}^{m}\}$ are
  deformations of the original theory introduced at the same time as the
  gauging of the global symmetries of the original symmetry is carried out. In
  this case they only have to be invariant under the global symmetries that
  have been gauged and not the stronger condition

\begin{equation}
\delta_{A} c^{\sharp} =0\, ,  
\end{equation}

\noindent
for any of them.

\item We can consider only the tensors $\{\vartheta_{I}{}^{A},Z^{Im},Z_{im}\}$
  are deformations of the original theory, whose definition includes the
  tensors $\{-d_{mIJ},d^{i}{}_{I}{}^{m}\}$. In this case, the latter must be
  invariant under the whole global symmetry group by hypothesis. The
  corresponding $Y_{A}{}^{\sharp}$ tensors are assumed to vanish identically,
  before they are contracted with the embedding tensor. This is the point of
  view that we have adopted here and it implies that there are only three sets
  of 8-form field strengths $\{L_{\sharp}\} =\{L_{A}{}^{I},L_{Im},L^{im}\}$
  and only three corresponding sets of 7-form potentials $\{E_{\sharp}\}
  =\{E_{A}{}^{I},E_{Im},E^{im}\}$ which are dual to the deformation tensors
  $\{\vartheta_{I}{}^{A},Z^{Im},Z_{im}\}$. In an action in which these tensors
  are generalized to spacetime-dependent fields, these dual potentials appear
  as Lagrange multipliers enforcing their constancy
  \cite{Bergshoeff:2007vb,Bergshoeff:2009ph}.

\end{enumerate}

We, thus, have to consider three constraints associated to gauge invariance

\begin{eqnarray}
\label{eq:gaugeinvarianceconstraint1}
\mathcal{Q}_{IJ}{}^{A} & \equiv & \vartheta_{I}{}^{B}Y_{BJ}{}^{A}\, , 
\hspace{1cm}
 Y_{BJ}{}^{A} \equiv  \delta_{B}\vartheta_{J}{}^{A}
= 
-T_{B}{}^{K}{}_{J}\vartheta_{K}{}^{A}+T_{B}{}^{A}{}_{C}\vartheta_{J}{}^{C}\, ,
\\
& & \nonumber \\
\label{eq:gaugeinvarianceconstraint2}
\mathcal{Q}_{I}{}^{Jm} & \equiv & \vartheta_{I}{}^{B}Y_{B}{}^{Jm}\, , 
\hspace{1cm}
Y_{B}{}^{Jm} \equiv \delta_{B}Z^{Jm}
=
T_{B}{}^{J}{}_{K}Z^{Km}+T_{B}{}^{m}{}_{n}Z^{Jn}\, ,
\\
& & \nonumber \\
\label{eq:gaugeinvarianceconstraint3}
\mathcal{Q}_{Iim} & \equiv & \vartheta_{I}{}^{B}Y_{Bim}\, , 
\hspace{1cm}
 Y_{Bim}  \equiv  \delta_{B}Z_{im}
=
-T_{B}{}^{j}{}_{i}Z_{jm}-T_{B}{}^{n}{}_{m}Z_{in}\, ,
\end{eqnarray}

\noindent
and two constraints associated to global invariance

\begin{eqnarray}
\label{eq:globalinvarianceconstraint1} 
\mathcal{Q}_{AmIJ} 
& \equiv & 
Y_{AmIJ} 
= 
-\delta_{A}d_{mIJ} 
=
T_{A}{}^{n}{}_{m}d_{nIJ}+2T_{A}^{K}{}_{(I}d_{|m|J)K}\, , 
\\
& & \nonumber \\
\label{eq:globalinvarianceconstraint2} 
\mathcal{Q}_{A}{}^{i}{}_{I}{}^{m} 
& \equiv & 
Y_{A}{}^{i}{}_{I}{}^{m} 
= 
\delta_{A}d^{i}{}_{I}{}^{m} 
=
T_{A}{}^{i}{}_{j}d^{j}{}_{I}{}^{m}-T_{A}{}^{J}{}_{I}d^{i}{}_{J}{}^{m}
+T_{A}{}^{m}{}_{n}d^{i}{}_{I}{}^{n}\, ,
\end{eqnarray}

\noindent
and the final form of the Bianchi identity for the 7-form field strengths is

\begin{equation}
\label{eq:Bianchigauged7}
\mathcal{D}K_{A}
= 
T_{A}{}^{K}{}_{J}F^{J}\tilde{F}_{K}  
+T_{A}{}^{m}{}_{n}\tilde{H}^{n}H_{m}
-\tfrac{1}{2}T_{A\, ij}G^{ij}
+Y_{AI}{}^{B}L_{B}{}^{I}+ Y_{A}{}^{Im}L_{Im} + Y_{Aim}L^{im}\, .
\end{equation}

The occurrence of these $Y_{A}{}^{\sharp}$ has to be confirmed by taking again
the covariant derivative of this Bianchi identity.

\subsection{9-form field strengths}
\label{sec-9-forms}

Taking the covariant derivative of both sides of the Bianchi identity
Eq.~(\ref{eq:Bianchigauged7}) we arrive to\footnote{By direct computation we
  have not found any constraint or $Y_{A}{}^{\sharp}$ tensor associated to
  either $d^{m}{}_{IJK}$ or $d^{mnp}$.}

\begin{equation}
  \begin{array}{rcl}
Y_{AI}{}^{B}\left[\mathcal{D}L_{B}{}^{I} +F^{I}K_{B}\right]
+Y_{A}{}^{Im}\left[ \mathcal{D}L_{Im} +\tilde{F}_{I}H_{m}\right]
+Y_{Aim}\left[\mathcal{D}L^{im}+G^{i}\tilde{H}^{m}\right]
& & \\
& & \\
+\mathcal{Q}_{AmIJ} \tilde{H}^{m}F^{IJ}
+\mathcal{Q}_{A}{}^{i}{}_{I}{}^{m}G_{i}F^{I}H_{m}
& = &
0\, .
\end{array}
\end{equation}

\noindent
Since we have assumed\footnote{Observe that, the alternative assumption is
  equally valid and can be made to work by including the 8-form field
  strengths $L^{mIJ},L_{i}{}^{I}{}_{m}$.}
$\mathcal{Q}_{AmIJ}=\mathcal{Q}_{A}{}^{i}{}_{I}{}^{m}=0$, we arrive to the
Bianchi identities

\begin{eqnarray}
\label{eq:Bianchigauged8-1}
\mathcal{D}L_{B}{}^{I}& = & -F^{I}K_{B} -W_{B}{}^{I\beta}M_{\beta}\, ,
\\
& & \nonumber \\
\label{eq:Bianchigauged8-2}
\mathcal{D}L_{Im} & = & -\tilde{F}_{I}H_{m} -W_{Im}{}^{\beta}M_{\beta}\, ,
\\
& & \nonumber \\
\label{eq:Bianchigauged8-3}
\mathcal{D}L^{im} & = & -G^{i}\tilde{H}^{m} -W^{im\beta}M_{\beta}\, ,
\end{eqnarray}

\noindent
where the $W_{\sharp}{}^{\beta}$ tensors are invariant tensors annihilated by
the $Y_{A}{}^{\sharp}$ ones

\begin{equation}
Y_{A}{}^{\sharp}W_{\sharp}{}^{\beta} =0\, .  
\end{equation}

As shown in Ref.~\cite{Hartong:2009vc} these tensors are nothing but the
derivatives of all the constraints satisfied by the deformation tensors
(labeled by $\beta$) with respect to the deformation tensors themselves. This
means that there are as many 9-form field strengths $M_{\beta}$ and
corresponding 8-form potentials $N_{\beta}$ as constraints
$\mathcal{Q}^{\beta}=0$. In a general action the top-form potentials
$N_{\beta}$ would occur as the Lagrange multipliers enforcing the constraints
$\mathcal{Q}^{\beta}=0$.

As usual, this can be confirmed by acting yet again with the covariant
derivative on the above three Bianchi identities. Let us first list all the
constraints we have met: 

\begin{enumerate}
\item First of all we have the gauge-invariance constraints 

  \begin{equation}
  \mathcal{Q}_{IJ}{}^{A}\, ,\,\,\, \mathcal{Q}_{I}{}^{Jm}\, ,\,\,\,
  \mathcal{Q}_{Iim}\, ,  
  \end{equation}

\noindent
  defined in
  Eqs.~(\ref{eq:gaugeinvarianceconstraint1})-(\ref{eq:gaugeinvarianceconstraint3}).

\item Secondly, we have the global-invariance constraints
  \begin{equation}
    \mathcal{Q}_{AmIJ}\, ,\,\,\, \mathcal{Q}_{A}{}^{i}{}_{I}{}^{m}\, , 
  \end{equation}

\noindent
defined in Eqs.~(\ref{eq:globalinvarianceconstraint1}) and
((\ref{eq:globalinvarianceconstraint2})).

\item Thirdly we have the orthogonality constraints between the three
  deformation tensors

  \begin{eqnarray}
  \mathcal{Q}^{mA} & \equiv & -Z^{Im}\vartheta_{I}{}^{A}\, ,
\\
& & \nonumber \\
\mathcal{Q}_{i}{}^{I} & \equiv &  Z_{im}Z^{Im}\, ,
\\
& & \nonumber \\
\mathcal{Q}_{mn} & \equiv &  Z_{im}Z^{i}{}_{n}\, .
  \end{eqnarray}

\item Next, we have the constraints relating the gauge transformations to
  the $d$-tensors

  \begin{eqnarray}
  \mathcal{Q}_{I}{}^{J}{}_{K} & \equiv &  X_{(I}{}^{J}{}_{K)}
  -Z^{Jm}d_{mIK}\, ,
\\
& & \nonumber \\
  \mathcal{Q}_{I}{}^{m}{}_{n} & \equiv &  X_{I}{}^{m}{}_{n} +2d_{nIJ}Z^{Jm}
  +Z_{in}d^{i}{}_{I}{}^{m}\, ,
\\
& & \nonumber \\
  \mathcal{Q}_{Iij} & \equiv &  -X_{Iij} -2Z_{(i|m}d_{|j)I}{}^{m}\, ,
  \end{eqnarray}

\item Finally, we have the constraints that related the $d$-tensors amongst
  them via the massive deformations $Z$

\begin{eqnarray}
\mathcal{Q}^{imn} & \equiv &  d^{i}{}_{I}{}^{[m|}Z^{I|n]}
+Z^{i}{}_{p}d^{pmn}\, ,
\\
& & \nonumber \\
\mathcal{Q}_{IJ}{}^{mn} & \equiv &
\tfrac{1}{2}d^{i}{}_{(I|}{}^{m}d_{i|J)}{}^{n}
+d^{mnp}d_{pIJ} +3 d^{[m|}{}_{IJK}Z^{K|n]}\, ,
\\
& & \nonumber \\ 
\mathcal{Q}_{iIJK} & \equiv & Z_{im}d^{m}{}_{IJK} -d_{i(I|}{}^{m}d_{m|JK)}\, .
  \end{eqnarray}
\end{enumerate}

From Eq.~(\ref{eq:Bianchigauged8-1}) we get

\begin{eqnarray}
\frac{\partial \mathcal{Q}_{IJ}{}^{A}}{\partial \vartheta_{K}{}^{B}} 
\left[\mathcal{D}M^{IJ}{}_{A}+F^{I}L_{A}{}^{J} \right] 
+\frac{\partial \mathcal{Q}^{mA}}{\partial \vartheta_{K}{}^{B}}
\left[\mathcal{D}M_{mA}+ H_{m}K_{A}\right]
& & \nonumber \\
& & \nonumber \\
+\frac{\partial \mathcal{Q}_{I}{}^{J}{}_{K}}{\partial \vartheta_{K}{}^{B}}
\left[\mathcal{D}M^{I}{}_{J}{}^{K}+ F^{IK}\tilde{F}_{J}\right]
+\frac{\partial \mathcal{Q}_{I}{}^{m}{}_{n}}{\partial \vartheta_{K}{}^{B}}
\left[\mathcal{D}M^{I}{}_{m}{}^{n}+ F^{I}\tilde{H}^{m}H_{n}\right]
& & \nonumber \\
& & \nonumber \\
+\frac{\partial \mathcal{Q}_{Iij}}{\partial \vartheta_{K}{}^{B}}
\left[\mathcal{D}M^{Iij}+ F^{I}G^{ij}\right]
& = & 0\, .
\end{eqnarray}

From Eqs.~(\ref{eq:Bianchigauged8-2}) and (\ref{eq:Bianchigauged8-3}) we get
very similar equations which guarantee the consistency of the whole
construction of the tensor hierarchy that we have carried out in this section.

\section{Gauge-invariant action for the 1-, 2- and 3-forms}
\label{sec-action}

The Bianchi identities of the full tensor hierarchy give rise to the equations
of motion of the electric fields of the theory upon use of the duality
relations (\textit{on-duality-shell}). For field strengths of the 6-,5-,
4-forms they are given by

\begin{equation}
K_{A}
=
-\star j_{A}^{(\sigma)}\, ,
\hspace{1cm}
\tilde{F}_{I}
=
\mathcal{M}_{IJ}\star F^{J}\, ,
\hspace{1cm}
\tilde{H}^{m}
= 
\star \mathcal{M}^{mn}H_{n}\, .
\end{equation}

\noindent
For the field strengths of the magnetic 3-forms they are given by 

\begin{equation}
G_{a} = R_{a}\, ,  
\end{equation}

\noindent
where $R_{a}$ has been defined in Eq.~(\ref{eq:Radef}).  Finally, the field
strength of the 7-forms is, according to
Refs.~\cite{Bergshoeff:2009ph,Hartong:2009vc}, dual to the derivatives of the
gauge-invariant scalar potential with respect to the deformation parameters,
denoted collectively by $c^{\sharp}$

\begin{equation}
L_{\sharp} = \star \frac{\partial V}{\partial c^{\sharp}}\, .
\end{equation}

This identity follows from the scalar equation of motion in presence of a
scalar potential together with the  condition

\begin{equation}
k_{A}{}^{x}\frac{\partial V}{\partial \phi^{x}} = 
Y_{A}{}^{\sharp}\frac{\partial V}{\partial c^{\sharp}}\, ,  
\end{equation}

\noindent
which implies, after multiplication by the embedding tensor
$\vartheta_{I}{}^{A}$, the gauge-invariance of the scalar potential.

In general, the equations of motion are combinations of different Bianchi
identities on-duality-shell. In order to determine the combinations that
correspond to the equations of motion we have to examine which combinations
of Bianchi identities satisfy the Noether identities associated to the gauge
invariances of the theory. 

To start with, we need to introduce some notation for the Bianchi
identities. This has been done in Appendix~\ref{app-Bianchisgauged}.  These
Bianchi identities are related by a hierarchy of identities that are obtained
by taking the covariant derivative of those with lower rank, as we have
shown. These identities of Bianchi identities are collected in
Appendix~\ref{app-identitiesofBianchiidentities}.

Now, let us assume that a standard gauge-invariant action for the 0-forms
$\mathcal{M}$ (or $\phi^{x}$), 1-forms $A^{I}$, 2-forms $B_{m}$ and electric
3-forms $C^{a}$ exists. This means that the Bianchi identities
$\mathcal{B}(\mathcal{Q}^{\beta}),\mathcal{B}(c^{\sharp})$ and
$\mathcal{B}(\mathcal{D}\mathcal{M}),\mathcal{B}(F^{I}),\mathcal{B}(H_{m}),
\mathcal{B}(G_{a})$ are satisfied, at least up to duality relations. The
kinetic terms of the electric fields are written in terms of the
gauge-invariant field strengths and this implies that the magnetic fields
$C_{a},\tilde{B}^{m}$ must necessarily occur in the action, albeit not as
dynamical fields: their equations of motion will be trivial on-duality-shell.

Under these assumptions, the identities satisfied by the non-trivial Bianchi
identities (\textit{i.e.}~those involving the magnetic field strengths) take
the simplified form\footnote{We have also ignored the identities whose rank,
  as differential forms, is higher than eight.}

\begin{eqnarray}
\mathcal{D}\mathcal{B}(H_{m}) -Z^{a}{}_{m}G_{a}
& = & 0\, ,
\\
& & \nonumber \\
\mathcal{D}\mathcal{B}(G_{a}) 
-Z_{am}\mathcal{B}(\tilde{H}^{m})\
& = & 0\, ,
\\
& & \nonumber \\
\mathcal{D}\mathcal{B}(\tilde{H}^{m}) 
+d^{a}{}_{I}{}^{m}\mathcal{B}(G_{a})F^{I}
+Z^{Im}\mathcal{B}(\tilde{F}_{I}) 
& = & 0\, ,
\\
& & \nonumber \\
\mathcal{D}\mathcal{B}(\tilde{F}_{I}) 
+2d_{mIJ}
\mathcal{B}(\tilde{H}^{m})F^{J}
-d^{a}{}_{I}{}^{m}\mathcal{B}(G_{a})H_{m}
+\vartheta_{I}{}^{A}\mathcal{B}(K_{A})
& = & 0\, .
\end{eqnarray}

If such an action exists, its invariance with respect to gauge transformations
with parameters $\sigma^{m},\sigma^{i},\sigma_{m},\sigma^{I}$ will imply that
the equations of motion satisfy, off-shell, associated Noether identities. Up
to the field equations of $\tilde{B}^{m}$ and $C_{a}$ which are assumed to be
satisfied up to dualities, they take the form

\begin{eqnarray}
\mathcal{D}\frac{\delta S}{\delta \tilde{B}^{m}} -Z^{a}{}_{m}\frac{\delta S}{\delta C_{a}}
& = & 
0\, ,
\\
& & \nonumber \\
\mathcal{D}\frac{\delta S}{\delta  C^{a}}
-Z_{am}\frac{\delta S}{\delta B_{m}} 
& = & 
0\, ,
\\
& & \nonumber \\
\mathcal{D}
\frac{\delta S}{\delta B_{m}} 
+Z^{Im}
\left[
\frac{\delta S}{\delta A^{I}} 
-d_{nIJ}A^{J}\frac{\delta S}{\delta B_{n}}
-\left(d^{a}{}_{I}{}^{n}B_{n}
-\tfrac{1}{3}d^{a}{}_{J}{}^{n}d_{nIK}A^{JK}\right)\frac{\delta S}{\delta C^{a}}
\right]
& & \nonumber \\
& & \nonumber \\
+d^{a}{}_{I}{}^{m}F^{I}\frac{\delta S}{\delta C^{a}}
& = & 
0\, ,
\\
& & \nonumber \\
\mathcal{D}
\left[
\frac{\delta S}{\delta A^{I}} 
-d_{nIJ}A^{J}\frac{\delta S}{\delta B_{n}}
-\left(d^{a}{}_{I}{}^{n}B_{n}
-\tfrac{1}{3}d^{a}{}_{J}{}^{n}d_{nIK}A^{JK}\right)\frac{\delta S}{\delta C^{a}}
\right]
& & \nonumber \\
& & \nonumber \\
+2d_{mIJ}F^{J}\frac{\delta S}{\delta B_{m}}
+d^{a}{}_{I}{}^{m}H_{m}\frac{\delta S}{\delta C^{a}}
+\vartheta_{I}{}^{A}k_{A}{}^{x}\frac{\delta S}{\delta \phi^{x}}
& = &
0\, .
\end{eqnarray}

Comparing directly with the above identities satisfied by the Bianchi
identities, we conclude that, up to dualities, the equations of motion of the
electric fields are related to the Bianchi identities of the magnetic field
strengths by

\begin{eqnarray}
k_{A}{}^{x}\frac{\delta S}{\delta \phi^{x}} 
& = &
\mathcal{B}(K_{A})\, ,
\\
& & \nonumber \\  
\frac{\delta S}{\delta A^{I}}
& = &
\mathcal{B}(\tilde{F}_{I})
+\left(
d^{a}{}_{I}{}^{m}B_{m} 
-\tfrac{1}{3}d^{a}{}_{J}{}^{m}d_{mIK}A^{JK} 
\right)\mathcal{B}(G_{a})
+d_{mIJ}A^{J}\mathcal{B}(\tilde{H}^{m})\, ,
\\
& & \nonumber \\  
\frac{\delta S}{\delta B_{m}}
& = & 
\mathcal{B}(\tilde{H}^{m})\, ,
\\
& & \nonumber \\  
\frac{\delta S}{\delta C^{a}}
& = &
\mathcal{B}(G_{a})\, .
\end{eqnarray}

This identification determines completely the field theory.
For instance, the equation of motion for the electric 3-forms $C^{a}$ must be

\begin{equation}
\label{eq:Caeom}
\frac{\delta S}{\delta C^{a}}
=
\mathcal{D}\left(
\Im\mathfrak{m}\mathcal{N}_{ab} \star G^{b}
+\Re\mathfrak{e}\mathcal{N}_{ab}  G^{b}
 \right)
+d_{aI}{}^{m} F^{I}H_{m} -Z_{am}\mathcal{M}^{mn}\star H_{n}\, ,
\end{equation}

\noindent
etc.

Can we write an action gauge-invariant action for the electric fields
$\phi^{x},A^{I},B_{m}$ and $C^{a}$ from which these equations of motion
follow, up to duality relations? We can follow the step-by-step procedure used
in Ref.~\cite{Hartong:2009vc} for the 5- and 6-dimensional cases. This
procedure consists in considering first an action $S^{(0)}$ containing only
the gauge-invariant kinetic terms for the all the electric potentials
$\phi^{x},A^{I},B_{m},C^{a}$ and start adding the necessary Chern-Simons terms
$S^{(1)},S^{(2)},\ldots$ to obtain the equations of motion of all the
potentials occurring in $S^{(0)}$ in order of decreasing rank:
$\tilde{B}^{m},C^{a},C_{a},B_{m},A^{I}$. At the first step it will be
necessary to introduce terms $S^{(1)}$ containing $\tilde{B}^{m}$ but no new
terms containing this potential will be introduced in the following steps. At
the second step we will introduce terms $S^{(2)}$ containing $C^{a}$ (but no
$\tilde{B}^{m}$) and in the following steps we will not introduce any more
terms containing it and so on and so forth.

We will not carry this procedure to the end because in eight dimensions the
number of Chern-Simons terms involving just 2- and 1-form potentials is huge
and its structure is very complicated. Nevertheless we are going to check that
everything works as expected for the potentials of highest rank
$\tilde{B}^{m},C^{a},C_{a}$ and we are going to find that only under certain
conditions the action we are looking for exists

Our starting point is, therefore, the action

\begin{equation}
\label{eq:d8gaugedaction0}
\begin{array}{rcl}
S^{(0)} 
& = & 
{\displaystyle\int} 
\left\{ 
-\star \mathbbm{1} R
+\tfrac{1}{2} \mathcal{G}_{xy}\mathcal{D}\phi^{x}\wedge \star \mathcal{D}\phi^{y}
+\frac{1}{2}\mathcal{M}_{IJ}F^{I}\wedge \star F^{J}
+\frac{1}{2}\mathcal{M}^{mn} H_{m}\wedge \star H_{n}
\right.
\\
& & \\
& & 
\left.
+\frac{1}{2} G^{a} \wedge R_{a}
-\star \mathbbm{1} V(\phi)
\right\} \, ,
\\
\end{array}
\end{equation}

\noindent 
where we have added a scalar potential $V(\phi)$. This action gives

\begin{equation}
\frac{\delta S^{(0)}}{\delta \tilde{B}^{m}}
=
-Z^{a}{}_{m}R_{a}\, .  
\end{equation}

\noindent
This equations should be trivial on-duality-shell and, therefore, we must add
to the action $S^{(0)}$

\begin{equation}
S^{(1)}
=
\int Z^{a}{}_{m}(G_{a}+\tfrac{1}{2}Z_{an}\tilde{B}^{n})\tilde{B}^{m}\, ,  
\end{equation}

\noindent
so that 

\begin{equation}
\frac{\delta (S^{(0)}+S^{(1)})}{\delta \tilde{B}^{m}}
=
-Z^{a}{}_{m}(R_{a}-G_{a})\, .  
\end{equation}

The equation for $C^{a}$ that follows from $S^{(0)}+S^{(1)}$ is

\begin{equation}
\frac{\delta (S^{(0)}+S^{(1)})}{\delta C^{a}}
=
-\mathcal{D}R_{a}
-Z_{am}\mathcal{M}^{mn}\star H_{n}\, ,  
\end{equation}

\noindent
and, comparing with Eq.~(\ref{eq:Caeom}), we see that the term $+d_{aI}{}^{m}
F^{I}H_{m}$ is missing and we must add a term of the form 

\begin{equation}
S^{(2)}
=
\int d_{aI}{}^{m} F^{I}\left(H_{m}-\tfrac{1}{2}Z_{bm}C^{b} \right) C^{a}\, , 
\end{equation}

Observe that $\tilde{B}^{m}$ does not appear in this term and its equation of
motion is, therefore, not modified by it.  However in this term or in any
other similar term the only part of $d_{aI}{}^{m}Z_{bm}$ that can occur is the
antisymmetric one $d_{[a|I}{}^{m}Z_{|b]m}$ while in the term $+d_{aI}{}^{m}
F^{I}H_{m}$ both the antisymmetric and the symmetric parts occur. The only way
in which we can get that term in the equations of motion is by requiring

\begin{equation}
\label{eq:demand1}
d_{(a|I}{}^{m}Z_{|b)m} = -\tfrac{1}{2} X_{I\, ab}=0\, .  
\end{equation}

\noindent
Under this assumption, which will also prove crucial to obtain the equations
of motion of other fields, the equation of motion of $C^{a}$ is
Eq.~(\ref{eq:Caeom}), as we wanted.

The equation of the magnetic potential $C_{a}$, which should be trivial
on-duality-shell which follows from the action we have constructed is

\begin{equation}
\frac{\delta (S^{(0)}+S^{(1)}+S^{(2)})}{\delta C^{a}}
=
Z^{a}{}_{m}
\left[
\mathcal{M}^{mn}\star H_{n} -\mathcal{D}\tilde{B}^{m} + d_{bI}{}^{m}F^{I}C^{b}
\right]\, .
\end{equation}

\noindent
The last two terms belong to the field strength $\tilde{H}^{m}$ and we need to
add 

\begin{equation}
S^{(3)}
=
\int 
\left\{
-\tfrac{1}{2}d^{b}{}_{I}{}^{m}Z^{a}{}_{m}F^{I}C_{b}C_{a}
-\left[
2d^{mnp}B_{n}(H_{p}-2Z_{ip}C^{i}-\tfrac{1}{2}\mathcal{D}B_{p})
+\Box \tilde{H}^{m}
\right]Z^{a}{}_{m}C_{a}
\right\}\, ,
\end{equation}

\noindent
where $\Box \tilde{H}^{m}$ is the part of the field strength $\tilde{H}^{m}$
that only contains 1-form potentials and their exterior derivatives.  Observe
that neither $\tilde{B}^{m}$ nor $C^{a}$ appear in this term and, therefore,
their equations of motion are not modified. Observe also that we are facing
here the same problem we faced in getting the equation of motion of $C^{a}$:
only $d^{[b|}{}_{I}{}^{m}Z^{|a]}{}_{m}$ can enter the action while the
equation of motion contains also the symmetric part. The solution to this
problem is the same: we demand

\begin{equation}
\label{eq:demand2}
d^{(a|}{}_{I}{}^{m}Z^{|b)}{}_{m} = -\tfrac{1}{2} X_{I}{}^{ab}=0\, .    
\end{equation}

Using Eqs.(\ref{eq:demand1}) and (\ref{eq:demand2}) The equation of motion of
$B_{m}$ that follows from the action $S^{(0)}+\cdots+S^{(3)}$ can be put in
the form

\begin{equation}
  \begin{array}{rcl}
{\displaystyle\frac{\delta (S^{(0)}+\cdots+S^{(3)})}{\delta B_{m}}}
& = &
-\left[
\mathcal{D}(\mathcal{M}^{mn}\star H_{n}) +d_{aI}{}^{m}F^{I}G^{a}
-d^{a}{}_{I}{}^{m}F^{I}R_{a} 
\right.
\\
& & \\
& & 
\left.
-d^{mnp}H_{np} 
-Z^{Im}\mathcal{M}_{IJ}\star F^{J}
\right]
-d^{mnp}B_{n}Z^{a}{}_{p}(R_{a}-G_{a})
\\
& & \\
& & 
-d^{mnp}Z^{a}{}_{p}d_{aI}{}^{q} B_{n}B_{q}(F^{I}-\tfrac{1}{2} Z^{Ir}B_{r})
\\
& & \\
& & 
+d_{aI}{}^{m}d^{a}{}_{J}{}^{n}F^{I}(F^{I}-\tfrac{1}{2} Z^{Ip}B_{p})B_{n}
\\
& & \\
& & 
+d_{aI}{}^{m}F^{I}\Box G^{a}+d^{a}{}_{I}{}^{[m|} Z^{I|n]} B_{n}\Box G_{a}
\\
& & \\
& & 
-d^{mnp}(H_{n}-Z_{in}C^{i})(H_{p}-Z_{jp}C^{j})\, .
\end{array}
\end{equation}

The expression in brackets in the r.h.s.~is identical to
$\mathcal{B}(\tilde{H}^{m})$ up to dualities and up to the term
$d^{m}{}_{IJK}F^{IJK}$. The next term vanishes on-duality-shell and the
remaining terms should be eliminated. Observe that in the terms that need to
be eliminated and introduced neither $\tilde{B}^{m}$ nor $C^{i}$ occur (they
only depend on $B_{m},A^{I}$ and their derivatives) and, therefore, their
equations of motion will not be modified.

\section{Conclusions}
\label{sec-conclusions}

Following the same procedure as in Refs.~\cite{Hartong:2009vc,Ortin:2015hya},
in this paper we have constructed the most general 8-dimensional theory with
gauge symnmetries and with at most two derivatives: field strengths (up to
6-forms), all the Bianchi identities and duality relations satisfied by all
the field strengths (up to the 9-forms\footnote{These identities are, of
  course, just formal, but they encode the gauge transformations of the 8-form
  potentials.}), and the equations of motion of the fundamental fields. We
have shown that they are characterized by a small number of invariant tensors
($d$-tensor, embedding tensor $\vartheta$ and massive deformations $Z$) that
satisfy certain constraints that relate them among themselves and to the
structure constants and generators of the global symmetry group, which has to
act on the $n_{3}$ 3-form potentials of the theory as a subgroup of
Sp$(2n_{3},\mathbb{R})$.

We have found that the Bianchi identities satisfied by the 7-form field
strengths (dual to the generalized Noether-Gaillard-Zumino current) have the
general form predicted in Ref.~\cite{Bandos:2016smv}, although in this case it is very
difficult to find the explicit form of the 7-form field strengths.

We have constructed an action from which one can derive all the equations of
motion except for those of the 1-form potentials because identifying the terms
that only contain 1-forms becomes extremely complicated and time-consuming.

This general result can be applied to any 8-dimensional theory with a given
field content, $d$-tensors defining Chern-Simons interactions and global
symmetry group, such as maximal $d=8$ supergravity. In a forthcoming
publication we will solve the constraints satisfied by the deformation tensors
($d$-tensor, embedding tensor $\vartheta$ and massive deformations $Z$)
searching for a 1-parameter family of different SO$(3)$ gaugings of that
theory.

\section*{Acknowledgments}

This work has been supported in part by the Spanish Ministry of Science and
Education grant FPA2012-35043-C02-01, the Centro de Excelencia Severo Ochoa
Program grant SEV-2012-0249, and the Spanish Consolider-Ingenio 2010 program
CPAN CSD2007-00042.  The work of OLA was further supported by a scholarship of
the Ecuadorian Secretary of Science, Technology and Innovation.  TO wishes to
thank M.M.~Fern\'andez for her permanent support.

\appendix
\section{Summary of relations for the generic ungauged, massless Abelian $d=8$ theory}
\label{app-genericmassless}

\subsection{Field strengths}

\begin{eqnarray}
F^{I} 
& = &
d A^{I}\, .
\\
& & \nonumber \\
H_{m} 
& = &
dB_{m} - d_{mIJ}F^{I}A^{J}\, ,
\\
& & \nonumber \\
G^{i} 
& = &
dC^{i} +d^{i}{}_{I}{}^{m}F^{I}B_{m} 
-\tfrac{1}{3} d^{i}{}_{I}{}^{m}d_{mJK} A^{I}F^{J}A^{K}\, ,
\\
& & \nonumber \\
\tilde{H}^{m}
& = &
d\tilde{B}^{m} 
+d^{i}{}_{I}{}^{m} C_{i}F^{I}  
+d^{mnp}B_{n}(H_{p}+\Delta H_{p})
+\tfrac{1}{12}d^{i}{}_{I}{}^{m}d_{iJ}{}^{n} A^{IJ}\Delta H_{n}\, ,
\\
& & \nonumber \\
\tilde{F}_{I}
& = &
d\tilde{A}_{I} +2d_{mIJ}A^{J}(\tilde{H}_{m}-\tfrac{1}{2}\Delta\tilde{H}_{m})
-\left( 
d^{i}{}_{I}{}^{m}B_{m} -\tfrac{1}{3}d^{i}{}_{J}{}^{m}d_{mIK}A^{JK}
\right)
(G_{i}-\tfrac{1}{2}\Delta G_{i})
\nonumber
\\
& & \nonumber \\
& & 
-\tfrac{1}{3}
\left(
d^{i}{}_{I}{}^{m}d_{mJK} -d^{i}{}_{K}{}^{m}d_{mIJ}
\right) 
F^{J}A^{K}C_{i}
-d^{mnp}d_{mIJ}A^{J}B_{n}H_{p}
\nonumber 
\\
& & \nonumber \\
& & 
+\tfrac{1}{24}
\left(
d^{i}{}_{K}{}^{m}d_{iL}{}^{n}d_{mIJ}
+2d^{i}{}_{[I|}{}^{m}d_{i|K]}{}^{n}d_{mJL}
\right)
F^{J}A^{KL}B_{n}
+\tfrac{1}{24} d^{i}{}_{J}{}^{m} d_{iK}{}^{n} d_{mIL}A^{JKL}dB_{n}
\nonumber 
\\
& & \nonumber \\
& & 
-\tfrac{1}{180} d^{i}{}_{L}{}^{n}d_{iQ}{}^{m}d_{mIJ} d_{nPK} A^{JKLQ}F^{P}\, ,
\end{eqnarray}

\subsection{Bianchi identities}

\begin{eqnarray}
dF^{I} 
& = &
0\, ,
\\
& & \nonumber \\
dH_{m} 
& = &
-d_{mIJ}F^{IJ}\, ,
\\
& & \nonumber \\
dG^{i} 
& = &
d^{i}{}_{I}{}^{m}F^{I}H_{m}\, ,
\\
& & \nonumber \\
d \tilde{H}^{m} 
& = &
d^{i}{}_{I}{}^{m}G_{i}F^{I} + d^{mnp}H_{np}\, ,
\\
& & \nonumber \\
d \tilde{F}_{I}
& = &
2d_{mIJ}F^{J}\tilde{H}^{m} +d_{iI}{}^{m}G^{i}H_{m}\, ,
\\
& & \nonumber \\
dK_{A}
& = &
T_{A}{}^{I}{}_{J}F^{J}\tilde{F}_{I}
+T_{A}{}^{m}{}_{n}\tilde{H}^{n}H_{m}
-\tfrac{1}{2}T_{Aij}G^{ij}\, .
\end{eqnarray}

\subsection{Duality relations}

\begin{eqnarray}
\star G^{i}
&  = &
\Omega^{ij}\mathcal{W}_{jk}G^{k}\, ,
\hspace{.5cm}
\mbox{or}
\hspace{.5cm}
G_{a}^{+} = -\mathcal{N}^{*}_{ab}G^{b\, +}\, , 
\\
& & \nonumber \\
\star \tilde{H}^{m}
& = &
\mathcal{M}^{mn}H_{n}\, ,
\\
& & \nonumber \\
\star \tilde{F}_{I}
& = &
-\mathcal{M}_{IJ}F^{J}\, ,
\\
& & \nonumber \\
\star K_{A}
& = &
-j_{A}^{(\sigma)}\, ,
\\
& & \nonumber \\
\star L_{\sharp}
& = &
-\frac{\partial V}{\partial c^{\sharp}}
\end{eqnarray}

\section{Summary of relations for the gauged theory}
\label{app-gauged}

\subsection{Field strengths}

\begin{eqnarray}
F^{I} 
& = &
dA^{I} -\tfrac{1}{2}X_{J}{}^{I}{}_{K}A^{JK} +Z^{I\,  m}B_{m}\, ,
\\
& & \nonumber \\
H_{m} 
& = &
\mathcal{D}B_{m} -d_{mIJ}dA^{I}A^{J} +\tfrac{1}{3}d_{mMI}X_{J}{}^{M}{}_{K}A^{IJK}
+Z_{im} C^{i}\, ,
\\
& & \nonumber \\
G^{i} 
& = &
\mathcal{D}C^{i} 
+d^{i}{}_{I}{}^{n}
\left[
F^{I}B_{n} -\tfrac{1}{2}Z^{I\, p}B_{n}B_{p}
+\tfrac{1}{3}d_{nJK}dA^{J}A^{KI}
\right.
\nonumber \\
& & \nonumber \\
& & 
\left.
+\tfrac{1}{12}d_{nMJ}X_{K}{}^{M}{}_{L} A^{IJKL}
\right]
-Z^{i}{}_{m}\tilde{H}^{m}\, ,
\\
& & \nonumber \\
\tilde{H}^{m}
& = &
\mathcal{D}\tilde{B}^{m} -d_{iI}{}^{m}F^{I}C^{i}
+d^{mnp}B_{n}\left(H_{p}+\Delta H_{p}-2Z_{ip}C^{i}\right)
\nonumber \\
& & \nonumber \\
& & 
+d^{m}{}_{IJK}dA^{I}dA^{J}A^{K}
\nonumber \\
& & \nonumber \\
& & 
+
\left(
\tfrac{1}{12}d_{iJ}{}^{m}d^{i}{}_{K}{}^{n}d_{nIL} 
-\tfrac{3}{4} d^{m}{}_{IJM} X_{K}{}^{M}{}_{L}
\right)
dA^{I} A^{JKL}
\nonumber \\
& & \nonumber \\
& & 
+
\left(
\tfrac{3}{20} d^{m}{}_{NPM}X_{I}{}^{N}{}_{J}
-\tfrac{1}{60} d_{iM}{}^{m}d^{i}_{I}{}^{n}d_{nPJ}
\right) 
X_{K}{}^{P}{}_{L} A^{IJKLM}
\nonumber \\
& & \nonumber \\
& & 
+Z^{Im}\tilde{A}_{I}\, ,
\end{eqnarray}

\subsection{Bianchi identities}
\label{app-Bianchisgauged}

The Bianchi identities satisfied by the field strengths of the gauged theory
are $\mathcal{B}(\cdot)=0$ where

\begin{eqnarray}
\mathcal{B}(L_{A}{}^{I})
& = &
-\left[
\mathcal{D}L_{A}{}^{I} +F^{I}K_{A} +W^{I}{}_{A}{}^{\beta}M_{\beta}
\right]\, ,
\\
& & \nonumber \\
\mathcal{B}(L_{Im})
& = &
-\left[
\mathcal{D}L_{Im} +\tilde{F}_{I}H_{m} +W_{Im}{}^{\beta}M_{\beta}
\right]\, ,
\\
& & \nonumber \\
\mathcal{B}(L^{im})
& = &
-\left[
\mathcal{D}L^{im} +G^{i}\tilde{H}^{m} +W^{im\, \beta}M_{\beta}
\right]\, ,
\\
& & \nonumber \\
\mathcal{B}(K_{A})
& = &
\mathcal{D}K_{A}
-T_{A}{}^{I}{}_{J}F^{J}\tilde{F}_{I} -
T_{A}{}^{m}{}_{n}\tilde{H}^{n}H_{m}
+\tfrac{1}{2}T_{A\, ij}G^{ij} -Y_{A}{}^{\sharp}L_{\sharp}\, ,
\\
& & \nonumber \\
\mathcal{B}(\tilde{F}_{I})
& = &
-\left[
\mathcal{D}\tilde{F}_{I}-2d_{mIJ} F^{J}\tilde{H}^{m}
-d_{iI}{}^{m}G^{i}H_{m}
+2d^{m}{}_{IJK}F^{JK}H_{m}+\vartheta_{I}{}^{A}K_{A}
\right]\, ,
\\
& & \nonumber \\
\mathcal{B}(\tilde{H}^{m})
& = &
-\left[
\mathcal{D} \tilde{H}^{m} 
+d_{iI}{}^{m}F^{I}G^{i} -d^{mnp}H_{np} -d^{m}{}_{IJK}F^{IJK} 
-Z^{Im}\tilde{F}_{I}
\right]\, ,
\\
& & \nonumber \\
\mathcal{B}(G_{i})
& = &
-\left[
\mathcal{D}G_{i} 
-d_{iI}{}^{m} F^{I}H_{m} +Z_{im}\tilde{H}^{m}
\right]\, ,
\\
& & \nonumber \\
\mathcal{B}(H_{m})
& = &
-\left[
\mathcal{D}H_{m}
+d_{mIJ}F^{IJ}-Z_{im}G^{i}
\right]\, , 
\\
& & \nonumber \\
\mathcal{B}(F^{I})
& = &
-\left[
\mathcal{D}F^{I} - Z^{Im}H_{m}
\right]\, ,
\\
& & \nonumber \\
\mathcal{B}(\mathcal{D}\mathcal{M})
& = &
-\left[
\mathcal{D}\mathcal{D}\mathcal{M}+F^{I}\vartheta_{I}{}^{A}\delta_{A}\mathcal{M}
\right]\, ,
\\
& & \nonumber \\
\mathcal{B}(c^{\sharp})
& = &
\mathcal{D}c^{\sharp}\, ,
\\
& & \nonumber \\
\mathcal{B}(\mathcal{Q}^{\beta})
& = &
\mathcal{Q}^{\beta}\, .
\end{eqnarray}

Here $\sharp$ labels the deformation parameters and $\beta$ the constraints,
as discussed in Sections~\ref{sec-8-forms} and~\ref{sec-9-forms}.

\subsection{Identities of Bianchi identities}
\label{app-identitiesofBianchiidentities}

\begin{eqnarray}
\mathcal{D}\mathcal{B}(F^{I}) -Z^{Im}\mathcal{B}(H_{m}) 
& = & 0\, ,
\\
& & \nonumber \\
\mathcal{D}\mathcal{B}(H_{m})-2d_{mIJ}F^{I}\mathcal{B}(F^{J}) +Z_{im}G^{i}
& = & 0\, ,
\\
& & \nonumber \\
\mathcal{D}\mathcal{B}(G_{i}) 
-d_{iI}{}^{m}
\left[\mathcal{B}(H_{m})F^{I}+H_{m}\mathcal{B}(F^{I})\right]
-Z_{im}\mathcal{B}(\tilde{H}^{m})\
& = & 0\, ,
\\
& & \nonumber \\
\mathcal{D}\mathcal{B}(\tilde{H}^{m}) 
-d_{iI}{}^{m}\left[\mathcal{B}(G^{i})F^{I}+G^{i}\mathcal{B}(F^{I}) \right]
+2d^{mnp}\mathcal{B}(H_{n})H_{p}
& & \nonumber \\
& & \nonumber \\
-3d^{m}{}_{IJK}\mathcal{B}(F^{I})F^{JK}
+Z^{Im}\mathcal{B}(\tilde{F}_{I}) 
& = & 0\, ,
\\
& & \nonumber \\
\mathcal{D}\mathcal{B}(\tilde{F}_{I}) 
+2d_{mIJ}
\left[\mathcal{B}(\tilde{H}^{m})F^{J}+\tilde{H}^{m}\mathcal{B}(F^{J})\right]
& & \nonumber \\
& & \nonumber \\
+d_{iI}{}^{m}\left[\mathcal{B}(G^{i})H_{m}+G^{i}\mathcal{B}(H_{m}) \right]
& & \nonumber \\
& & \nonumber \\
-3d^{m}{}_{IJK}
\left[2\mathcal{B}(F^{J})F^{K}H_{m} +F^{JK}\mathcal{B}(H_{m}) \right]
+\vartheta_{I}{}^{A}\mathcal{B}(K_{A})
& = & 0\, ,
\\
& & \nonumber \\
\mathcal{D}\mathcal{B}(K_{A})
+T_{A}{}^{I}{}_{J}\left[\mathcal{B}(F^{J})\tilde{F}_{I}
+F^{J}\mathcal{B}(\tilde{F}_{I})\right]
& & \nonumber \\
& & \nonumber \\
+T_{A}{}^{m}{}_{n}\left[\mathcal{B}(\tilde{H}^{n})H_{m}
+\tilde{H}^{n}\mathcal{B}(H_{m}) \right]
+T_{A\, ij}G^{i}\mathcal{B}(G^{j})
+Y_{A}{}^{\sharp}\mathcal{B}(L_{\sharp}) 
& = & 0\, .
\end{eqnarray}



\begin{thebibliography}{99}

\bibitem{Salam:1989ihk}
A.~Salam and E.~Sezgin,
``Supergravities in Diverse Dimensions'' (Vols.~1 and 2)
World Scientific, Singapore (1989).
\doi{10.1142/0277}


\bibitem{Cordaro:1998tx}
F.~Cordaro, P.~Fr\'e, L.~Gualtieri, P.~Termonia and M.~Trigiante,
``N=8 gaugings revisited: An Exhaustive classification,''
Nucl.\ Phys.\ B {\bf 532} (1998) 245.
\doi{10.1016/S0550-3213(98)00449-0}.
[\hepth{9804056}].

\bibitem{Nicolai:2000sc}
H.~Nicolai and H.~Samtleben,
``Maximal gauged supergravity in three-dimensions,''
Phys.\ Rev.\ Lett.\  {\bf 86} (2001) 1686.
\doi{10.1103/PhysRevLett.86.1686}.
[\hepth{0010076}].

\bibitem{Nicolai:2001sv}
H.~Nicolai and H.~Samtleben,
``Compact and noncompact gauged maximal supergravities in three-dimensions
,''    
JHEP {\bf 0104} (2001) 022.
\doi{10.1088/1126-6708/2001/04/022}
[\hepth{0103032}].

\bibitem{deWit:2002vt}
B.~de Wit, H.~Samtleben and M.~Trigiante,
``On Lagrangians and gaugings of maximal supergravities,''
Nucl.\ Phys.\ B {\bf 655} (2003) 93.
\doi{10.1016/S0550-3213(03)00059-2}
[\hepth{0212239}].

\bibitem{deWit:2005ub}
B.~de Wit, H.~Samtleben and M.~Trigiante,
``Magnetic charges in local field theory,''
JHEP {\bf 0509} (2005) 016.
\doi{10.1088/1126-6708/2005/09/016}.
[\hepth{0507289}].


\bibitem{Meessen:1998qm}
P.~Meessen and T.~Ort\'{\i}n,
``An Sl(2,Z) multiplet of nine-dimensional type II supergravity theories,''
Nucl.\ Phys.\ B {\bf 541} (1999) 195.
\doi{10.1016/S0550-3213(98)00780-9}.
[\hepth{9806120}].

\bibitem{AlonsoAlberca:2000gh}
N.~Alonso-Alberca, P.~Meessen and T.~Ort\'{\i}n,
``{\sl An Sl(3,Z) multiplet of eight-dimensional type II supergravity theories and the gauged supergravity inside,}''
Nucl.\ Phys.\ B {\bf 602} (2001) 329.
\doi{10.1016/S0550-3213(01)00110-9}.
[\hepth{0012032}].

\bibitem{Gheerardyn:2001jj}
J.~Gheerardyn and P.~Meessen,
``Supersymmetry of massive D = 9 supergravity,''
Phys.\ Lett.\ B {\bf 525} (2002) 322.
\doi{10.1016/S0370-2693(01)01429-0}.
[\hepth{0111130}].

\bibitem{AlonsoAlberca:2002tb}
N.~Alonso-Alberca and T.~Ort\'{\i}n,
``{\sl Gauged / massive supergravities in diverse dimensions,}''
Nucl.\ Phys.\ B {\bf 651} (2003) 263.
\doi{10.1016/S0550-3213(02)01125-2}
[\hepth{0210011}].

\bibitem{AlonsoAlberca:2003jq}
N.~Alonso-Alberca, E.~Bergshoeff, U.~Gran, R.~Linares, T.~Ort\'{\i}n and D.~Roest,
``{\sl Domain walls of D = 8 gauged supergravities and their D = 11 origin,}''
JHEP {\bf 0306} (2003) 038.
\doi{10.1088/1126-6708/2003/06/038}.
[\hepth{0303113}].

\bibitem{Bergshoeff:2003ri}
E.~Bergshoeff, U.~Gran, R.~Linares, M.~Nielsen, T.~Ort\'{\i}n and D.~Roest,
``{\sl The Bianchi classification of maximal D = 8 gauged supergravities,}''
Class.\ Quant.\ Grav.\  {\bf 20} (2003) 3997.
\doi{10.1088/0264-9381/20/18/310}.
[\hepth{0306179}].

\bibitem{Scherk:1979zr}
J.~Scherk and J.~H.~Schwarz,
``{\sl How to Get Masses from Extra Dimensions,}''
Nucl.\ Phys.\ B {\bf 153} (1979) 61.
\doi{10.1016/0550-3213(79)90592-3}

\bibitem{Salam:1984ft}
A.~Salam and E.~Sezgin,
``d = 8 Supergravity,''
Nucl.\ Phys.\ B {\bf 258} (1985) 284.
\doi{10.1016/0550-3213(85)90613-3}.

\bibitem{Ortin:2015hya}
T.~Ort\'{\i}n,
``Gravity and Strings'', 2nd edition, 
Cambridge University Press, 2015.

\bibitem{Puigdomenech:2008kia}
D.~Puigdomenech,
``{\sl Embedding tensor approach to maximal D = 8 supergravity,}''
\href{http://inspirehep.net/record/1285294/files/Thesis-2008-Puigdomenech.pdf}{Master
  Thesis, Groningen U., 2008.}

\bibitem{Hohm:2015xna}
O.~Hohm and Y.~N.~Wang,
``{\sl Tensor Hierarchy and Generalized Cartan Calculus in SL(3)$\times$SL(2) Exceptional Field Theory,}''
JHEP {\bf 1504} (2015) 050.
\doi{10.1007/JHEP04(2015)050}
[\arxiv{1501.01600} [hep-th]].

\bibitem{Hartong:2009vc}
J.~Hartong and T.~Ort\'{\i}n,
``{\sl Tensor Hierarchies of 5- and 6-Dimensional Field Theories,}''
JHEP {\bf 0909} (2009) 039.
\doi{10.1088/1126-6708/2009/09/039}.
[\arxiv{0906.4043} [hep-th]].

\bibitem{Bandos:2016smv}
I.~A.~Bandos and T.~Ort\'{\i}n,
``On the dualization of scalars into (d-2)-forms in supergravity. Momentum maps, R-symmetry and gauged supergravit
y,''
\arxiv{1605.05559} [hep-th].

\bibitem{Gaillard:1981rj}
M.~K.~Gaillard and B.~Zumino,
``Duality Rotations for Interacting Fields,''
Nucl.\ Phys.\ B {\bf 193} (1981) 221.
\doi{10.1016/0550-3213(81)90527-7}.

\bibitem{Andrianopoli:1996cm}
L.~Andrianopoli, M.~Bertolini, A.~Ceresole, R.~D'Auria, S.~Ferrara, 
P.~Fr\'e and T.~Magri,
``{\sl N=2 supergravity and N=2 superYang-Mills theory on general scalar manifolds: Symplectic covariance, gaugings and the momentum map,}''
J.\ Geom.\ Phys.\  {\bf 23} (1997) 111.
\doi{10.1016/S0393-0440(97)00002-8}.
[\hepth{9605032}].

\bibitem{Dall'Agata:2012bb}
G.~Dall'Agata, G.~Inverso and M.~Trigiante,
``{\sl Evidence for a family of SO(8) gauged supergravity theories,}''
Phys.\ Rev.\ Lett.\  {\bf 109} (2012) 201301.
\doi{10.1103/PhysRevLett.109.201301}.
[\arxiv{1209.0760} [hep-th]].

\bibitem{Bergshoeff:2009ph}
E.~A.~Bergshoeff, J.~Hartong, O.~Hohm, M.~H\"ubscher and T.~Ort\'{\i}n,
``{\sl Gauge Theories, Duality Relations and the Tensor Hierarchy},''
JHEP {\bf 0904} (2009) 123.
\doi{10.1088/1126-6708/2009/04/123}.
[\arxiv{0901.2054} [hep-th]].

\bibitem{Bergshoeff:2007vb}
E.~A.~Bergshoeff, J.~Gomis, T.~A.~Nutma and D.~Roest,
``{\sl Kac-Moody Spectrum of (Half-)Maximal Supergravities},''
JHEP {\bf 0802} (2008) 069.
\doi{10.1088/1126-6708/2008/02/069}.
[\arxiv{0711.2035} [hep-th]].

\bibitem{deWit:2005hv}
B.~de Wit and H.~Samtleben,
``Gauged maximal supergravities and hierarchies of nonAbelian vector-tens
or systems,''
Fortsch.\ Phys.\  {\bf 53} (2005) 442.
\doi{10.1002/prop.200510202}.
[\hepth{0501243}].

\bibitem{deWit:2008ta}
B.~de Wit, H.~Nicolai and H.~Samtleben,
``Gauged Supergravities, Tensor Hierarchies, and M-Theory,''
JHEP {\bf 0802} (2008) 044
\doi{10.1088/1126-6708/2008/02/044}.
[\arxiv{0801.1294} [hep-th]].

\bibitem{deWit:2009zv}
B.~de Wit and M.~van Zalk,
``Supergravity and M-Theory,''
Gen.\ Rel.\ Grav.\  {\bf 41} (2009) 757
\doi{10.1007/s10714-008-0751-0}.
[\arxiv{0901.4519} [hep-th]].

\bibitem{Bergshoeff:2007ef}
E.~Bergshoeff, H.~Samtleben and E.~Sezgin,
``{\sl The Gaugings of Maximal D=6 Supergravity},''
JHEP {\bf 0803} (2008) 068.
\doi{10.1088/1126-6708/2008/03/068}.
[\arxiv{0712.4277} [hep-th]].

\bibitem{kn:OLA}
``{\sl A 1-parameter family of SO$(3)$-gauged maximal $d=8$ supergravities},''
\'O.~Lasso Andino and T.~Ort\'{\i}n, to appear.


\end{thebibliography}
\end{document}